\documentclass[12pt,preprint]{aastex}

\usepackage{epsfig}
\shorttitle{Multiple Star HD 193322}
\shortauthors{ten Brummelaar et al.}

\begin{document}

%%\received{}
%%\accepted{}

\title{An Interferometric and Spectroscopic Analysis \\ 
 of the Multiple Star System HD~193322} 

\author{
Theo A. ten Brummelaar\altaffilmark{1},
David P. O'Brien\altaffilmark{2},
Brian D. Mason\altaffilmark{3}, \\
Christopher D. Farrington\altaffilmark{1},
Alexander W. Fullerton\altaffilmark{4,5},
Douglas R. Gies\altaffilmark{2}, \\
Erika D. Grundstrom\altaffilmark{6,7},
William I. Hartkopf\altaffilmark{3}, 
Rachel A. Matson\altaffilmark{2},
Harold A. McAlister\altaffilmark{2},
M. Virginia McSwain\altaffilmark{7,8},
Lewis C. Roberts, Jr.\altaffilmark{9},
Gail H. Schaefer\altaffilmark{1},
Sergio Sim\'{o}n-D\'{i}az\altaffilmark{10}, 
Judit Sturmann\altaffilmark{1},
Laszlo Sturmann\altaffilmark{1},
Nils H. Turner\altaffilmark{1},
and Stephen J. Williams\altaffilmark{2,11}}

\altaffiltext{1}{Center for High Angular Resolution Astronomy, Georgia State University, Mt.\ Wilson, CA 91023; 
theo@chara-array.org, farrington@chara-array.org, schaefer@chara-array.org, nils@chara-array.org}

\altaffiltext{2}{Center for High Angular Resolution Astronomy, 
Department of Physics and Astronomy, Georgia State University, P.O. Box 4106, Atlanta, GA 30302-4106; 
obrien@chara.gsu.edu, gies@chara.gsu.edu, rmatson@chara.gsu.edu, hal@chara.gsu.edu, swilliams@chara.gsu.edu}

\altaffiltext{3}{U.\ S.\ Naval Observatory, 3450 Massachusetts Avenue, NW, Washington, DC 20392-5420; 
bdm@usno.navy.mil, wih@usno.navy.mil}

\altaffiltext{4}{Space Telescope Science Institute, 3700 San Martin Drive, Baltimore, MD 21218;
fullerton@stsci.edu}

\altaffiltext{5}{Based on observations obtained at the Canada-France-Hawaii Telescope (CFHT) 
which is operated by the National Research Council of Canada, the Institut National des Sciences 
de l'Univers of the Centre National de la Recherche Scientifique of France,  and the University of Hawaii.}

\altaffiltext{6}{Physics and Astronomy Department, Vanderbilt University, 6301 Stevenson Center, 
Nashville, TN 37235; erika.grundstrom@vanderbilt.edu} 

\altaffiltext{7}{Visiting Astronomer, Kitt Peak National Observatory,
National Optical Astronomy Observatory, operated by the Association
of Universities for Research in Astronomy, Inc., under contract with
the National Science Foundation.}

\altaffiltext{8}{Department of Physics, Lehigh University, 
16 Memorial Drive East, Bethlehem, PA 18015; mcswain@lehigh.edu}

\altaffiltext{9}{Jet Propulsion Laboratory, California Institute of Technology, 
Adaptive Optics and Astronomical Instrumentation Group, 
4800 Oak Grove Drive, Pasadena CA 91109; lewis.c.roberts@jpl.nasa.gov}

\altaffiltext{10}{Instituto de Astrof\'{i}sica de Canarias, E-38200 La Laguna, Tenerife, Spain; 
Departamento de Astrof\'{i}sica, Universidad de La Laguna, E-38205, La Laguna, Tenerife, Spain; 
ssimon@iac.es}

\altaffiltext{11}{Guest investigator, Dominion Astrophysical Observatory, 
Herzberg Institute of Astrophysics, National Research Council of Canada.}

\slugcomment{Version 3 - Accepted by AJ}

%%\paperid{}

%%%%%%%%%%%%%%%%%%%%%%%%%%%%%%%%%%%%%%%%%%%%%%%%%%%%%%%%%%%%%%

\begin{abstract}

The star HD 193322 is a remarkable multiple system 
of massive stars that lies at the heart of the cluster 
Collinder 419.  Here we report on new spectroscopic 
observations and radial velocities of the narrow-lined component Ab1
that we use to determine its orbital motion around a close companion Ab2
($P = 312$ d) and around a distant third star Aa ($P = 35$ y).
We have also obtained long baseline interferometry of the 
target in the $K^\prime$-band with the CHARA Array that we use
in two ways.  First, we combine published speckle 
interferometric measurements with CHARA separated fringe packet 
measurements to improve the visual orbit for the wide Aa,Ab binary.
Second, we use measurements of the fringe packet 
from Aa to calibrate the visibility of the fringes 
of the Ab1,Ab2 binary, and we analyze these fringe visibilities 
to determine the visual orbit of the close system. 
The two most massive stars, Aa and Ab1, have masses of 
approximately 21 and $23 M_\odot$, respectively, and their 
spectral line broadening indicates that they represent extremes 
of fast and slow projected rotational velocity, respectively. 
\end{abstract}


%%%%%%%%%%%%%%%%%%%%%%%%%%%%%%%%%%%%%%%%%%%%%%%%%%%%%%%%%%%%%%%

\setcounter{footnote}{11}

\section{Introduction}                              % Section 1

Massive O-type stars are usually found with one or more nearby companion
(Mason et al.\ 1998, 2009).  Most of these luminous stars are very distant, and 
consequently, we generally only detect their very nearby companions through 
their Doppler shifts or very distant companions that are angularly resolved.
We must rely on high angular resolution observations of the few nearby 
cases to detect those elusive, mid-range separation binary stars. 
One of the most revealing examples is HD~193322 (O9~V:((n)); Walborn 1972), 
the central star in the sparse open cluster Collinder~419.  
The distance to the cluster is $741 \pm 36$~pc according to the recent
study by Roberts et al.\ (2010).  The star's complex multiplicity became
apparent with the discovery of a companion Ab through speckle interferometry 
observations by McAlister et al.\ (1987).  They designated the system 
as CHARA~96~Aa (McAlister et al.\ 1989), and subsequent speckle measurements
detected its orbital motion (Hartkopf et al.\ 1993; Hartkopf 2010).  
The Aa,Ab pair was also recently resolved through the 
technique of lucky imaging by
Ma\'{i}z Apell\'{a}niz (2010).  The composite optical spectrum 
is dominated by a relatively narrow-lined component Ab1, and Fullerton (1990)
discovered significant radial variations in this component indicative
of a spectroscopic binary.  The first spectroscopic orbit for Ab1 was 
presented by McKibben et al.\ (1998), who determined an orbital period
of 311~d.  In addition to the close Ab1,Ab2 spectroscopic pair and the 
speckle Aa,Ab pair, there is another wider companion B at an angular separation 
of 2.68 arcsec (Turner et al.\ 2008).  Components C and D are more distant
companions that also occupy the central region of Collinder 419 (Roberts
et al.\ 2010), and it is uncertain whether they are orbitally bound to
the central multiple system.  A mobile diagram presenting the known components of 
the system is illustrated in Figure~1.  The long orbital period estimate
for A,B pair is based upon the probable masses (see Table 8 below), distance,
and the assumption that the projected separation is the semimajor axis.  

\begin{figure}
\begin{center}
{\includegraphics[height=12cm]{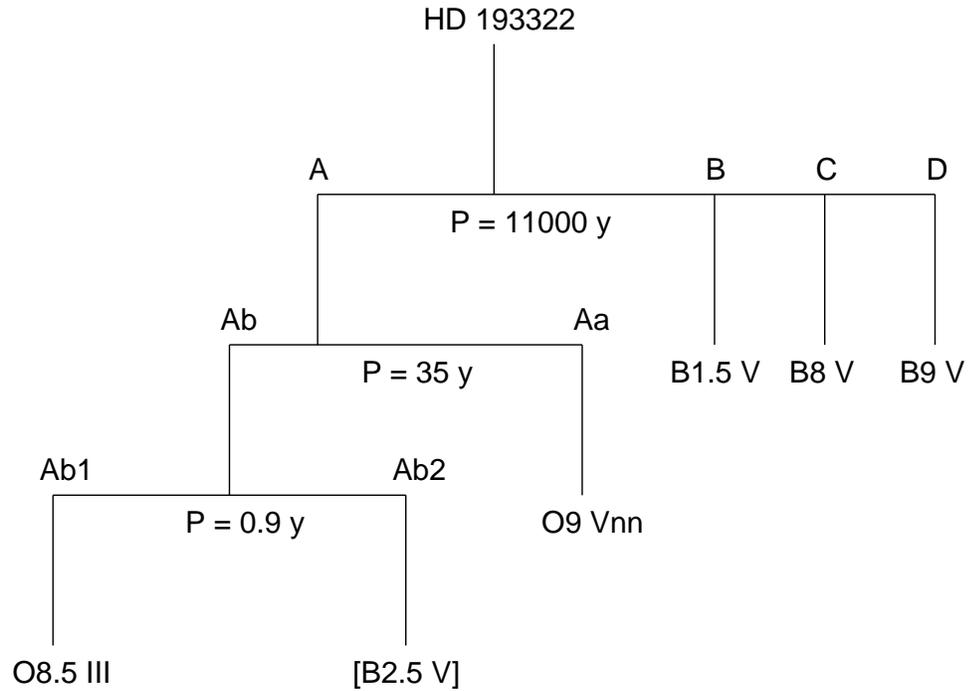}}
\end{center}
\caption{A mobile diagram of the components of the multiple star HD~193322.
The spectral classification is given under each stellar component.
The classification for Ab2 is enclosed in brackets to emphasize its
uncertainty (based upon its relative flux contribution; see \S3 and \S5).
The classifications for C and D are from Roberts et al.\ (2010). 
The period estimate for A,B is based upon the projected separation
(Mason et al.\ 1998).  
\label{fig1}}
\end{figure}

\placefigure{fig1}     % Figure 1 - Mobile diagram

Measurements of the orbital motions of the stars in this system offer us the 
means to estimate the masses of the components.  We have continued our 
interferometric and spectroscopic monitoring of the system over the 
last decade, and here we present a progress report on the orbits, 
mass estimates, and spectral properties of the component stars.
Combined speckle and interferometric observations of the motion 
of the Aa,Ab pair are used in \S2 to derive a preliminary orbit 
for the wide system.  We present in \S3 new long baseline interferometric 
measurements of the Ab1,Ab2 pair that are calibrated 
using the visibility of the Aa companion.  In \S4, we describe a 
diverse collection of spectroscopic observations that we use to 
derive a revised orbit for the narrow-lined Ab1 component in the 
close pair.  In \S5, we apply a Doppler tomography algorithm to a subset 
of the blue spectra to extract the spectra of the components. 
Finally in \S6, we discuss the masses and other properties of 
the components of the system. 

%%%%%%%%%%%%%%%%%%%%%%%%%%%%%%%%%%%%%%%%%%%%%%%%%%%%%%%%%%%%%%%

\section{Visual Orbit of the Wide System}           % Section 2

The orbital motion of the Aa,Ab pair has been followed since its 
discovery through continued speckle interferometry observations made
mainly with the Mayall 4~m telescope at Kitt Peak National Observatory  
(McAlister et al.\ 1989, 1993; Mason et al.\ 1998, 2009).
The date, position angle $\theta$, and separation $\rho$ of these
previously published observations are collected in Table~1 for convenience. 
While outliers exist, for this magnitude difference $\Delta m$ and $\rho$ 
regime, the errors from speckle interferometry measures are approximately 
$0\fdg5$ in position angle and $0.5\%$ in separation. 
We have also measured the relative motion through 
optical long baseline interferometry (OLBI) with the GSU CHARA Array 
at Mount Wilson Observatory (ten Brummelaar et al.\ 2005). 
These separation and position angle measurements are determined by 
measuring the fringe packet separation, when possible along two pairs of
baselines with approximately orthogonal directions projected onto the 
sky (the separated fringe packet or SFP method; Farrington et al.\ 2010).   
This separation is determined by fringe fitting in order to avoid shifts 
caused by overlap.  Other methods, like fits of the fringe envelope, 
may suffer if one fringe packet overlaps the secondary 
lobe of the other and causes the center of the fringe envelope to move.  
Like lunar occultation measurements, 
a single baseline measurement provides a separation in one direction only.  
Each of these measurements defines a line in astrometric space, and 
observations at several projected angles are required to define fully 
the position of the secondary.  The location of the secondary is defined 
as the point with the minimum total rms distance from these lines, as 
weighted by the variance of the fringe separations.  Formal errors are 
calculated using a method analogous to a $\chi^2$ analysis, and the errors 
for $\rho$ and $\theta$ are defined as the distance change required to 
increase the weighted rms by 1.0 in $\chi^2$.
These data have fairly low signal to noise and for many
epochs we have data from only a single baseline with a varying 
position angle from diurnal motion, so there is more scatter in the resulting
astrometry than that in the speckle data. 
% REFERENCE TO ONLINE OIFITS FILE HERE.
All the speckle and CHARA measurements are collected in Table~1 and are also
available as part of the online materials in an OIFITS file.. 

\placetable{tab1}      % Table 1 - visual orbital measurements

We made a new orbital solution for the wide pair using the combined set 
of speckle and long baseline interferometric observations.  
Note that we did not include the measurement from 
Ma\'{i}z Apell\'{a}niz (2010) using lucky imaging with the AstraLux
instrument because of its relatively large error. 
All the measurements were initially assigned equal weight, 
but in the orbital fitting process we identified three
discrepant points with large residuals that we subsequently 
zero-weighted in the final fit (those dates are marked in Table~1).  
The orbit was determined using the grid search method described by 
Hartkopf et al.\ (1989). The orbital elements are listed in Table~2 and 
the appearance of the visual orbit on the sky 
is shown in Figure~2.  The original speckle data set covers about $12\%$ 
and the recent speckle and SFP data cover about $15\%$ of the 
35~y orbit.  Note that we ignored making corrections for the 
center of light motion of the Ab binary, because the largest 
astrometric shifts are expected to be small, $\approx 0.5$~mas.

\begin{figure}
\begin{center}
{\includegraphics[width=15cm]{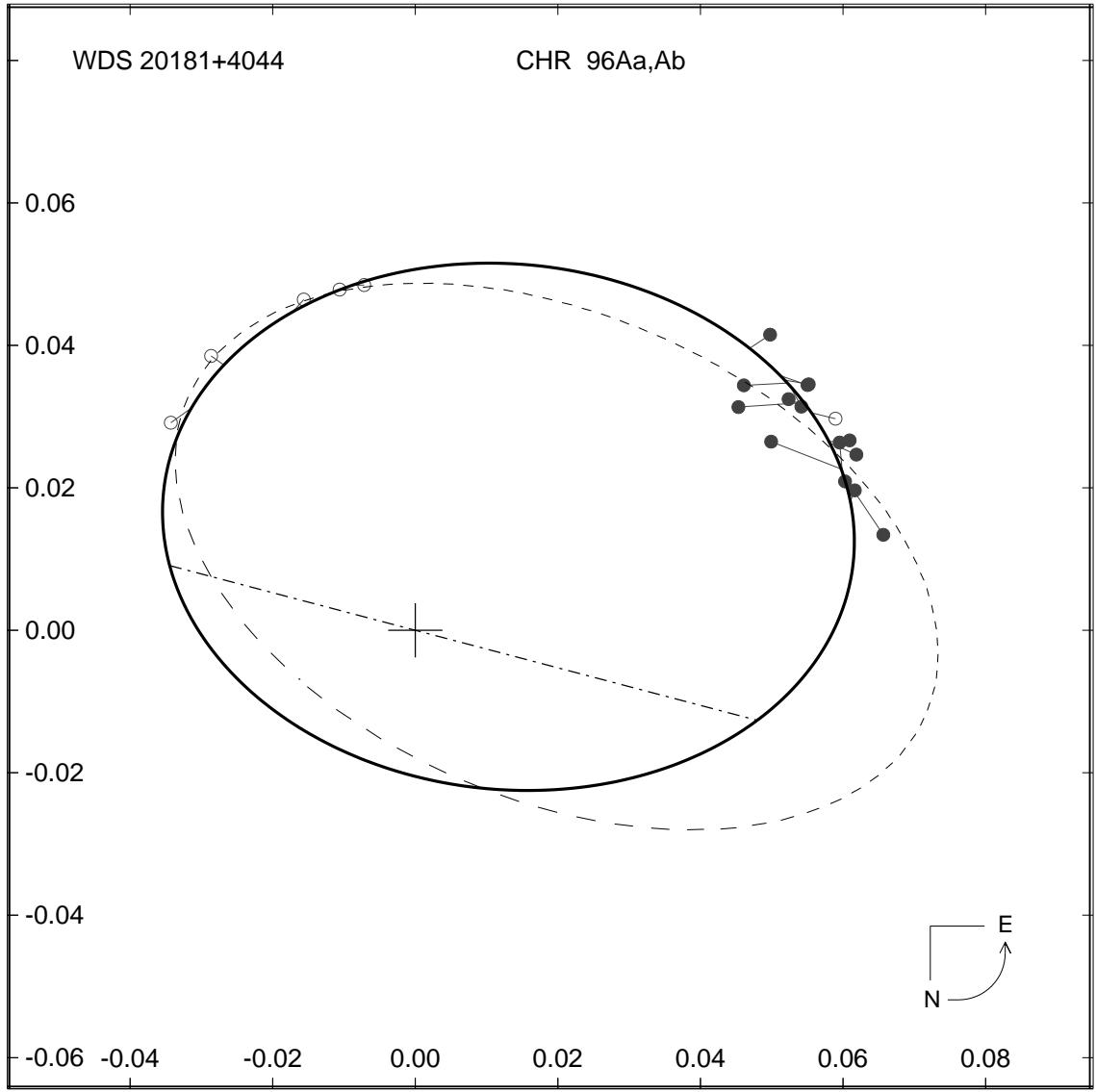}}
\end{center}
\caption{The astrometric orbit of the Aa,Ab pair (in units of arcsec). 
The dashed curve represents the first solution (Hartkopf et al.\ 1993) 
while the solid curve is the new solution (Table~2). 
The dot-dash line shows the line of the nodes. 
The filled circles represent the CHARA Array separated fringe packet results 
and open circles represent the speckle observations. 
Each measurement is connected by a line segment to the
calculated position for the time of observation. 
Note that north is down and east to the right in this figure, 
and the directional arc in the lower right corner shows the 
counter-clockwise sense of orbital motion.\label{fig2}}
\end{figure}

\placefigure{fig2}     % Figure 2 - wide orbit

\placetable{tab2}      % Table 2 - visual orbital elements

%%%%%%%%%%%%%%%%%%%%%%%%%%%%%%%%%%%%%%%%%%%%%%%%%%%%%%%%%%%%%%%

\section{Visual Orbit of the Close System}          % Section 3

The interferometric fringe patterns of the close Ab1,Ab2 pair 
of stars overlap even with the longest baselines available at
the CHARA Array, so we cannot use the separated fringe packet 
method to measure the relative separation.  
However, the interference of the two fringe patterns
of the inner pair causes a modulation of their combined visibility 
amplitude with changes in the projected baseline separation, and 
we can use this modulation to estimate the binary separation 
projected along the baseline position angle in the sky  
(Hummel et al.\ 1998; Boden et al.\ 2000; Raghavan et al.\ 2009). 
The calibration of visibility is aided when the signal of a nearby star
produces a separated fringe packet that can be used to calibrate the 
visibility of the central binary.  The details of this method 
are outlined by O'Brien et al.\ (2011).  

We obtained 195 observations of HD~193322 over 24 nights between 2005 and 2010
using the CHARA Array Classic beam combiner (ten Brummelaar et al.\ 2005).  
The observations were made with the 
near-IR $K^\prime$ filter and a variety of different baselines.
These measurements were series of approximately 200 recorded fringe scans  
sampled at a frequency of 150~Hz.
The scans were reduced by standard techniques 
(ten Brummelaar et al.\ 2005), and a subset of 100 scans with the best
S/N were selected.  We fit these scans with 
fringe patterns for each of the calibrator (Aa) and target (Ab) 
(identified according to the orbital solution from \S2 and the
position angle of the specific baseline; see details of the 
procedure in O'Brien et al.\ 2011).  We retained only those 
visibility measurements for which the fractional difference between 
a first estimate and final mean were less than $20\%$, and these 
best-case values were used to form the mean visibility for each component.  
Note that we used the same data set as that for the wide pair (\S2), but due to
the stronger selection criteria not all data sets yielded useful visibility amplitudes.
We next determined the ratio of the mean visibilities of the target and calibrator. 
This observed ratio is related to the ratio of the individual 
visibilities for the target and calibrator by
\begin{equation}
V_{\rm Ab} / V_{\rm Aa} = (F_{\rm Aa} / F_{\rm Ab})~ V_{\rm Ab,o} / V_{\rm Aa,o}
\end{equation}
where $F_{\rm Aa} / F_{\rm Ab}$ is the monochromatic flux 
ratio in the $K^\prime$-band.  The angular diameter of the calibrator Aa 
is small enough that $V_{\rm Aa}\approx 1.0$ for our observations, 
but we estimated the single-star visibilities of each component based upon 
the projected baseline of observation and the predicted angular diameters. 
A value for the flux ratio of 
$F_{\rm Aa} / F_{\rm Ab} = 10^{-0.4 \triangle m_{\rm wide}} = 0.92$ was adopted 
based upon a fit of the observed ratios and a binary model (see below), 
and this parameter essentially normalizes the target visibility so that the 
upper distribution of the visibilities has a mean of one.  Note that we 
expected that the ratio would be $F_{\rm Aa} / F_{\rm Ab} \geq 1$ based 
on the speckle orbit assignments (\S2), but we suspect that this difference
is probably insignificant given the uncertainties in the component flux fractions. 
The results for the Ab pair are given in Table~3 that lists 
the heliocentric Julian date of observation, the corresponding 
orbital phase in the Ab1,Ab2 orbit (\S4), the projected baseline $B_p$ and 
position angle $\psi_p$ of observation, the calibrated visibility and 
its associated error, and the observed minus calculated difference $O-C$
in visibility from the adopted model fit. 

\placetable{tab3}      % Table 3 - visibility ratio measurements

The modulation of the visibility ratio depends on the known 
projected baseline length and position angle and 
the effective wavelength of the $K^\prime$ system, 
plus the unknown projected binary separation and magnitude differences, 
$\triangle m_{\rm close} = -2.5 \log (F_{\rm Ab2} / F_{\rm Ab1})$ and 
$\triangle m_{\rm wide} = -2.5 \log (F_{\rm Aa} / F_{\rm Ab})$. 
The latter magnitude difference normalizes the visibility 
according to the relation given above, while the former 
magnitude difference sets the amplitude of the visibility 
modulation with baseline (Raghavan et al.\ 2009; O'Brien et al.\ 2011). 
Following the example of O'Brien et al.\ (2011), we explored  
the orbital parameters of the Ab1,Ab2 pair by creating a set 
of model visibilities for each of the observed times and 
baseline parameters and then forming the $\chi^2$ statistic
for the differences between the observed and model visibilities.
The solution is found by determining the orbital parameters 
and magnitude differences in a high resolution grid of values 
that minimize $\chi^2$.  For this application of the method, 
we set the orbital period and epoch from the spectroscopic 
elements for the circular orbit of Ab1,Ab2 (\S4) and then made 
a grid search for the best fit values of the angular semimajor axis $a$, 
inclination $i$, and longitude of the ascending node $\Omega$, 
plus the magnitude differences $\triangle m_{\rm close}$ and 
$\triangle m_{\rm wide}$.  

We found that the solutions always arrived at similar estimates
for the magnitude differences, so we set these magnitude differences
and performed a grid search over $i$ and $a$, the two parameters
of physical interest.  For each selection of $(i,a)$, we determined 
the best fit for the sky orientation parameter $\Omega$ over 
the full range of values in steps of $2^\circ$.  The resulting 
$\chi^2$ estimates are plotted in a gray-scale diagram in Figure~3
in the $(i,a)$ plane for grid increments of $\triangle i = 2^\circ$ and 
$\triangle a =0.05$~mas.  Here intensity is scaled between the lowest (black) 
and highest (white) $\chi^2$ over the grid.   If we assume a 
distance $d$ from the cluster fitting results of Roberts et al.\ (2010), 
then Kepler's Third Law relates the known period $P$, the total 
mass, and $a$ by 
\begin{equation}
(M({\rm Ab1}) + M({\rm Ab2}))/M_\odot = {{(a d)^3} \over {P^2}} = 
 \left({{a} \over {1.22~{\rm mas}}}\right)^3 ~\left({d\over {741~{\rm pc}}}\right)^3.
\end{equation}
Next, we can use the spectroscopic semiamplitude $K$ for 
component Ab1 (\S4, Table~6) to derive a relation for the mass of Ab2
as a function of $i$ and $a$, 
\begin{equation}
M({\rm Ab2}) = {{a^2} \over {P \sin i}} {{K (1-e^2)^{1/2}} \over {29.8~{\rm km~s}^{-1}}}
 = 0.455 M_\odot ~{a^2 \over {\sin i}} ~\left({d\over {741~{\rm pc}}}\right)^2. 
\end{equation}
Then we can find the mass of Ab1 from a relation for the mass ratio, 
\begin{equation}
{{M({\rm Ab1})} \over {M({\rm Ab2})}} =
  {{29.8~{\rm km~s}^{-1}} \over {K}} {{a d \sin i} \over {P \sqrt{1 - e^2}}} - 1
  = {{a \sin i} \over {0.819~{\rm mas}}}~\left({d\over {741~{\rm pc}}}\right) - 1.
\end{equation}
Thus, each point in the $(i,a)$ plane
is associated with specific masses $M({\rm Ab1})$ and $M({\rm Ab2})$, 
and we can use the relations above to construct loci of constant
primary and secondary mass in Figure~3 (shown by solid and dashed 
lines, respectively). 

\begin{figure}
\begin{center}
{\includegraphics[width=16cm]{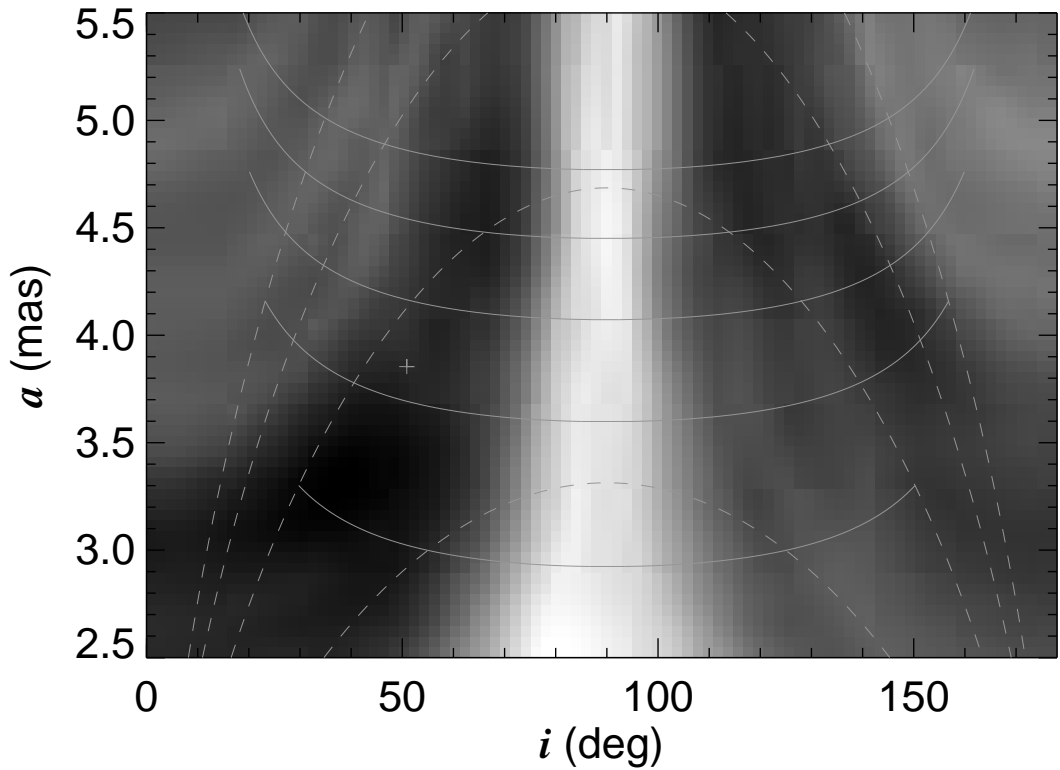}}
\end{center}
\caption{A gray-scale representation of the minimum $\chi^2$ over the 
full range in $\Omega$ for visibility models of the Ab1,Ab2 binary as a
function of orbital inclination $i$ and angular semimajor axis $a$.
Within this numerical grid of $(i,a)$, the minimum is $\chi^2=167$
(black) and the maximum is $\chi^2=529$ (white) for a sample of 
195 measurements and five fitting parameters.  The solid lines indicate
loci of constant $M$(Ab1) (10, 20, 30, 40, and $50 M_\odot$ from 
bottom to top), while the dashed lines represent loci of constant 
$M$(Ab2) (5, 10, 15, and $20 M_\odot$ from bottom to top), all 
for an assumed distance of 741~pc.  The location of the adopted 
solution is marked by a plus sign.\label{fig3}}
\end{figure}

\placefigure{fig3}     % Figure 3 - chi-squared of fits in (i,a) plane

Inspection of Figure~3 indicates that there are two broad 
valleys in the $(i,a)$ plane where the fits are relatively good,  
one with $i<90^\circ$ for counterclockwise motion in the sky and 
another with $i>90^\circ$ for clockwise motion.   Within these
valleys there are three locations with comparable minima, 
but all of these are associated with extreme masses:
$(i,a) = (66^\circ, 4.6~{\rm mas})$ ($\chi^2 = 210$) and 
$(i,a) = (118^\circ, 4.7~{\rm mas})$ ($\chi^2 = 215$) 
where the masses are too high and 
$(i,a) = (38^\circ, 3.3~{\rm mas})$ ($\chi^2 = 167$)  
where the masses are too low (see \S6 below).  
We think that the $i<90^\circ$ valley probably represents 
the best family of solutions, since the trends in $\chi^2$
are more or less continuous there as expected.  
Between $a=2.7$ and 5.0~mas, the valley floor never rises 
above a reduced chi-square of $\chi^2_\nu = 1.18$ 
(with 190 degrees of freedom, equal to 
195 measurements minus five fitting parameters).
Although a purely statistical assessment would restrict
the solution space to the valley region around 
$(i,a) = (38^\circ, 3.3~{\rm mas})$, the fact that the 
reduced chi-square is close to unity along the length of
the valley suggests that at this stage it is premature to rule 
out any of this solution space.  In \S6 below, we present several lines 
of argument that indicate that the actual solution lies
in the mid-range of this valley at $a=3.85$~mas ($\chi^2 = 225$), 
so we will tentatively adopt this value and 
present the associated solution for the other 
orbital parameters in Table~2, column~3.  
The errors associated with $i$, $a$, and $\Omega$  
reported in Table~2 correspond to their range over the
length of the valley from $a=2.7$ to 5.0~mas.  
Note that because the visibility oscillation 
depends on the absolute value of the projected separation, 
there is a $180^\circ$ ambiguity in our derived value of $\Omega$. 
We found that the best fit magnitude differences are
$\triangle m_{\rm close} = 2.11 \pm 0.06$ mag and 
$\triangle m_{\rm wide} = 0.086 \pm 0.012$ mag (Ab brighter than Aa). 
In order to show how well the model and observed visibilities agree, 
we plot the individual and calculated visibilities for this solution  
for each night in Figures~4 and 5, and we find that the fits are 
satisfactory for most of the nights. 

\begin{figure}
\begin{center}
{\includegraphics[angle=90,width=16cm]{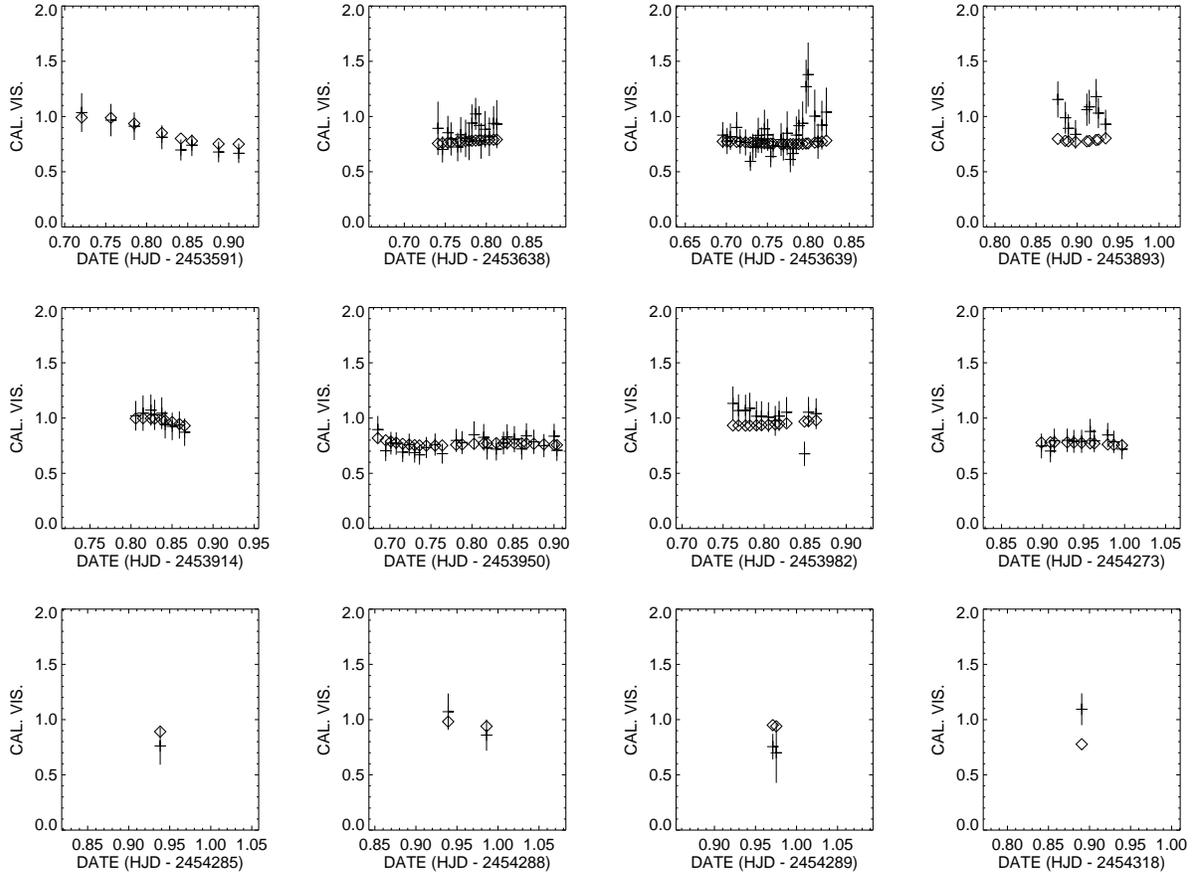}}
\end{center}
\caption{Plots of the calibrated (plus sign) and model (diamond) interferometric 
visibilities for component Ab1,Ab2 for each of the first 12 nights of 
observation with the CHARA Array.  
\label{fig4}}
\end{figure}

\placefigure{fig4}     % Figure 4 - visibilities first 12 nights

\begin{figure}
\begin{center}
{\includegraphics[angle=90,width=16cm]{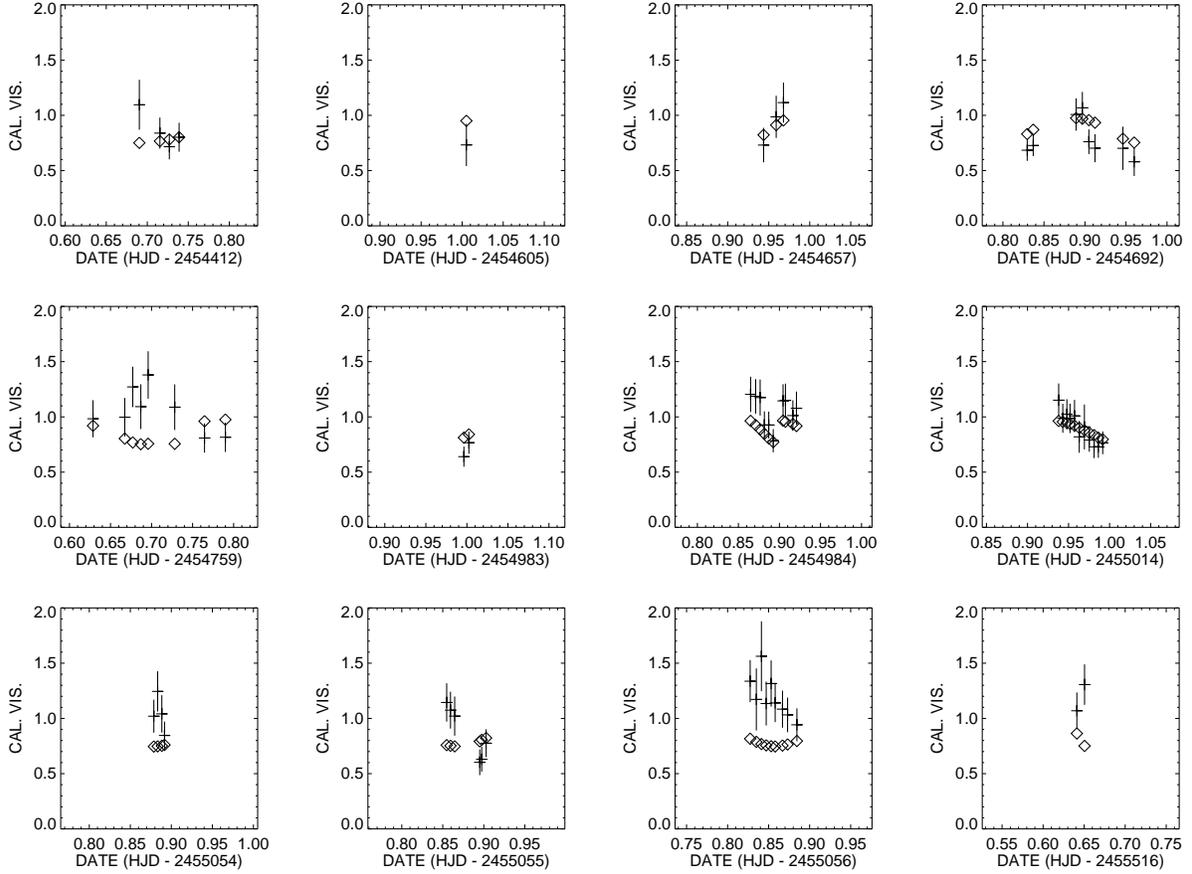}}
\end{center}
\caption{Plots of the calibrated (plus sign) and model (diamond) interferometric 
visibilities for component Ab1,Ab2 for each of the last 12 nights of observation 
with the CHARA Array. Note that the self-calibration method used for these data are extremely 
seeing dependent and this can cause large differences between the model and the data on some evenings. 
This is not unusual in interferometric data of low signal to noise.
\label{fig5}}
\end{figure}

\placefigure{fig5}     % Figure 5 - visibilities last 12 nights

The projection of the orbit on the sky for this solution  
is illustrated in Figure~6 where filled circles indicate the 
calculated positions at the times of observation.  The distribution 
of the observations in orbital phase appears to constrain the 
minor axis of the projected ellipse better than the major
axis.  With the minor axis fixed, the major axis will vary 
with inclination as $a({\rm minor}) \sim a \cos i$ or
$a \propto \sec i$, and this relation approximately describes
the position of the $\chi^2$ valley in Figure~3.  The orbital 
orientations of the wide and close orbits appear quite different
(compare Fig.~2 and Fig.~6), and the mutual inclination of the orbital
planes $\phi$ is given by Fekel (1981) as
\begin{equation}
\cos \phi = \cos i_{\rm close} \cos i_{\rm wide} \pm 
 \sin i_{\rm close} \sin i_{\rm wide} \cos (\Omega _{\rm close} - \Omega _{\rm wide})
\end{equation}
where the second term may assume either sign because of the $180^\circ$ 
ambiguity in the determination of $\Omega _{\rm close}$.  
The two solutions, $\phi = 38^\circ$ and $85^\circ$, indicate that
the two orbits are probably far from co-planar ($\phi = 0^\circ$). 

\begin{figure}
\begin{center}
{\includegraphics[width=15cm]{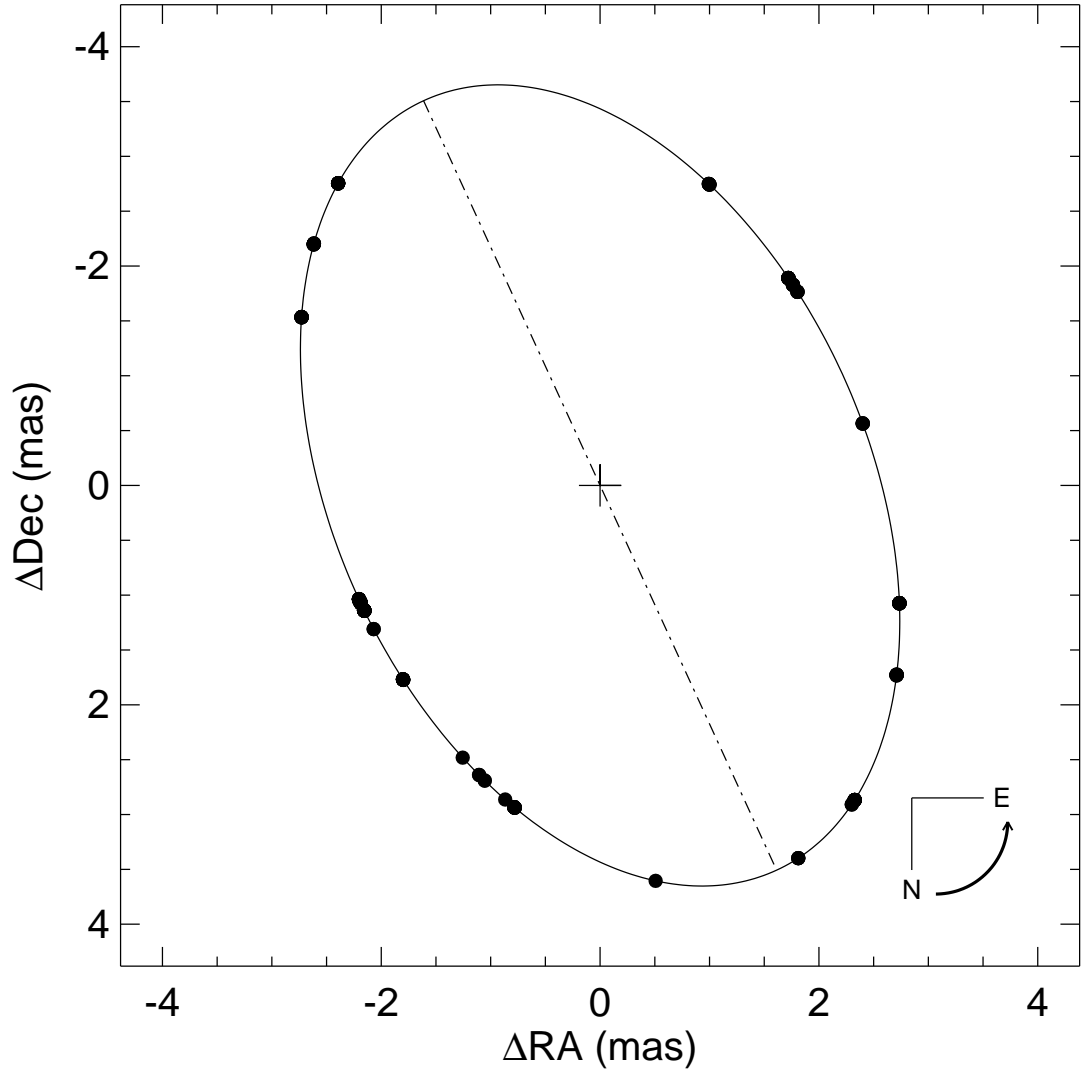}}
\end{center}
\caption{The astrometric orbit of the Ab1,Ab2 pair in the same format 
as Fig.~2 (but now in units of milli-arcsec = mas) based upon the  
CHARA Array visibility measurements.  Filled circles indicate the calculated 
positions at the times of observation.  
\label{fig6}}
\end{figure}

\placefigure{fig6}     % Figure 6 - close orbit

%%%%%%%%%%%%%%%%%%%%%%%%%%%%%%%%%%%%%%%%%%%%%%%%%%%%%%%%%%%%%%%

\section{Radial Velocities of Ab1}                  % Section 4

Spectroscopy potentially offers us the means to determine 
the masses and spectral properties of the components.  
However, because the stars are so close, conventional, ground-based 
spectroscopy records the flux of all three stars (usually plus 
component B at a separation of $2\farcs7$; Turner et al.\ 2008), 
and since their orbital Doppler shifts are comparable to the line widths, 
the resulting line blending problem is daunting.  Nevertheless, 
the spectral properties of the components are sufficiently 
different in this case that we may attempt radial velocity measurements. 
The appearance of the optical spectrum is dominated by a
narrow-lined component that corresponds to the primary star in the 
close orbit, Ab1 (McKibben et al.\ 1998).  The secondary in the close orbit, 
Ab2, is fainter and contributes little to the composite spectrum (\S5). 
Furthermore, there is a very broad-lined component that appears 
to follow the motion of Aa in the wide orbit (\S5).  
Because the lines of Aa are so broad and shallow, they 
essentially act to depress the continuum in the vicinity 
of the narrow lines of Ab1, and since the velocity range of 
Ab1 is smaller than the full width of the lines of Aa in general, 
the presence of the broad component has little influence on 
velocity measurements of Ab1 (but see a discussion of 
blending effects below).  Here we present radial velocities
for the Ab1 component and show that they represent the 
sum of orbital motions in both the close and wide systems. 

We collected 31 new spectra for measurement from sources 
that are summarized in Table~4.  The columns list a source 
number (for identification with the specific radial velocities 
listed in Table~5), date of observation(s), spectral range 
used in the measurement, the spectral resolving power, 
number of spectra made at that time, and the observatory,
telescope aperture, and spectrograph of origin.  We obtained 
most of these spectra in runs at the Kitt Peak National 
Observatory (KPNO) 0.9~m coud\'{e} feed and 4~m Mayall Telescopes,
the 3.6~m Canada-France-Hawaii Telescope, the 2.5~m Nordic Optical Telescope,
the Lowell Observatory 1.8~m Perkins Telescope, and the 
Herzberg Institute of Astrophysics, Dominion Astrophysical Observatory 
1.8~m Plaskett Telescope.  These were augmented with publicly 
available spectra from the archives of the University of Toledo 
Ritter Observatory 1.0~m telescope (Morrison et al.\ 1997), 
the Observatoire de Haute-Provence 1.9~m telescope and 
ELODIE spectrograph (Moultaka et al.\ 2004), and the Indo-U.S.\ Library 
of Coud\'{e} Feed Stellar Spectra (Valdes et al.\ 2004; 
made with the KPNO 0.9~m coud\'{e} feed telescope). 
All these spectra were reduced by standard techniques and 
transformed to a continuum normalized flux representation on 
a heliocentric, $\log$ wavelength grid.  Atmospheric telluric
lines were removed from the red spectra by division with a
pure atmospheric spectrum.  This was done by
creating a library of spectra from each run of a rapidly rotating 
A-star (usually $\zeta$~Aql), removing the broad stellar features from these, 
and then dividing each target spectrum by the modified atmospheric 
spectrum that most closely matched the target spectrum in a selected 
region dominated by atmospheric absorptions.

\placetable{tab4}      % Table 4 - sources of observations

The spectra form a diverse collection with a wide range in 
resolving power and wavelength coverage.  In order to measure 
radial velocities in a consistent way, we cross-correlated 
each of the spectra with a standard, model spectrum (rest frame)
from the grid of synthetic spectra from Lanz \& Hubeny (2003). 
From an initial inspection of the observations, we selected 
a model with Galactic abundances, effective temperature 
$T_{\rm eff} =34.8$~kK, gravity $\log g =4.0$, projected 
rotational velocity $V\sin i = 50$ km~s$^{-1}$, a wavelength 
dependent limb darkening coefficient from Wade \& Rucinski (1985), 
and an instrumental broadening appropriate for the specific  
observed spectrum.  The cross-correlations were generally 
made over the wavelength range given in Table~4, although in 
some cases regions with strong interstellar features were omitted. 
The resulting cross-correlation functions were always single-peaked, 
and we measured the radial velocity and its associated error 
using the method of Zucker (2003).  The results are presented 
in Table~5 that lists the heliocentric Julian date of mid-exposure, 
the corresponding orbital phases in the close and wide systems 
(see below), the measured radial velocity and its associated error,
a correction term for line blending effects,  
the observed minus calculated velocity residual from the fit  
(see below), and the observation source number from Table~4. 
Note that for completeness we have included in Table~5 velocities 
published earlier by McKibben et al.\ (1998; indicated by a 0 in the 
final column).  

\placetable{tab5}      % Table 5 - radial velocity measurements

Although the line blending effects from the spectral components 
of the other stars are generally small, they tend to bias the 
measurements towards the systemic velocity and lead to a slight 
underestimate of the orbital semiamplitude.  The radial velocity
offset caused by line blending will depend on the character and 
velocity shift of each component, the spectral features measured, 
and the spectral resolution of the observation.  In order to
make a simple correction for line blending effects we adopted 
the following procedure for each observation.  We first determined
model synthetic spectra for each stellar component (\S5) for 
the spectral range and instrumental broadening of the observation. 
The models for components Aa, Ab2, and B were co-added according
to the adopted fluxes and to the Doppler shifts for the time of observation. 
Then we formed a series of model spectra by adding in component 
Ab1 for a grid of assigned velocity offsets, and we measured 
the radial velocity in these composite spectra using the same 
cross-correlation method applied to the observations. 
This led to a relation between the actual and measured radial 
velocity for each observation, and we interpolated within this 
relation at the observed radial velocity to determine the 
offset correction for blending  
$\Delta V_{r}({\rm blend}) = V_r({\rm actual}) - V_r({\rm measured})$, 
which is given in Table~5, column~6.  The average 
of the absolute value of the offset correction for blending 
is small, 2.3 km~s$^{-1}$, but the individual offset corrections are larger 
for the lower resolution spectra where line blending is more severe. 

The velocities of component Ab1 depend on its orbital motion in the 
close binary plus the motion of the Ab1,Ab2 center of mass in the orbit 
of the Aa,Ab system.  Our first solutions for the orbital motion 
of the close binary clearly showed long term variations in the residuals 
that followed the motion predicted for Ab in the wide orbit. 
Thus, we fit the observed radial velocity variations as the 
sum of motions in the close and wide binaries.  This was done iteratively 
using the orbital fitting program of Morbey \& Brosterhus (1974). 
We first made a general fit of the velocities for the close system, 
and then we made a constrained fit of the velocity residuals, 
by fixing $P$, $e$ and $\omega$ from the visual orbit of Aa,Ab
to find the semiamplitude $K$ and epoch of periastron $T$ for the wide system.
The resulting solution of the long period orbit was then used to correct 
the observed velocities for motion in the wide orbit, and a new 
solution was found for the close orbit.  This procedure quickly 
converged to yield the orbital elements given in Table~6. 
Note that we assigned each measurement a weight proportional to
$\sigma (V_{r})^{-2}$ in making the fits, and we zero-weighted 
four measurements that had unusually large residuals from the final fit 
(dates indicated in Table~5 and shown as open circles in Fig.~7).
Table~6 lists the solutions both with and without application 
of the offset correction for line blending, and they are generally 
very similar except for the slightly larger semiamplitude $K$
that results when accounting for line blending.  Since the line 
blending problem is significant, we adopt the corrected velocity 
solutions that are illustrated in Figure~7 (close orbit)
and Figure~8 (wide orbit) and that form the basis for the 
residuals $O-C$ given in Table~5, column~7.  
We found that the eccentricity associated
with the close orbit is not statistically different from zero 
according to the criterion of Lucy \& Sweeney (1971), so we 
present circular elements in Table~6 (where the epoch $T$ 
is defined as the time of maximum radial velocity or, equivalently, 
the time of crossing the ascending node).  
The long orbital period of the wide system, $P = 35$~y, 
places HD~193322 among the top $1\%$ of known spectroscopic binaries 
with very long periods (Pourbaix et al.\ 2004). 

\begin{figure}
\begin{center}
\includegraphics[angle=90,width=15cm]{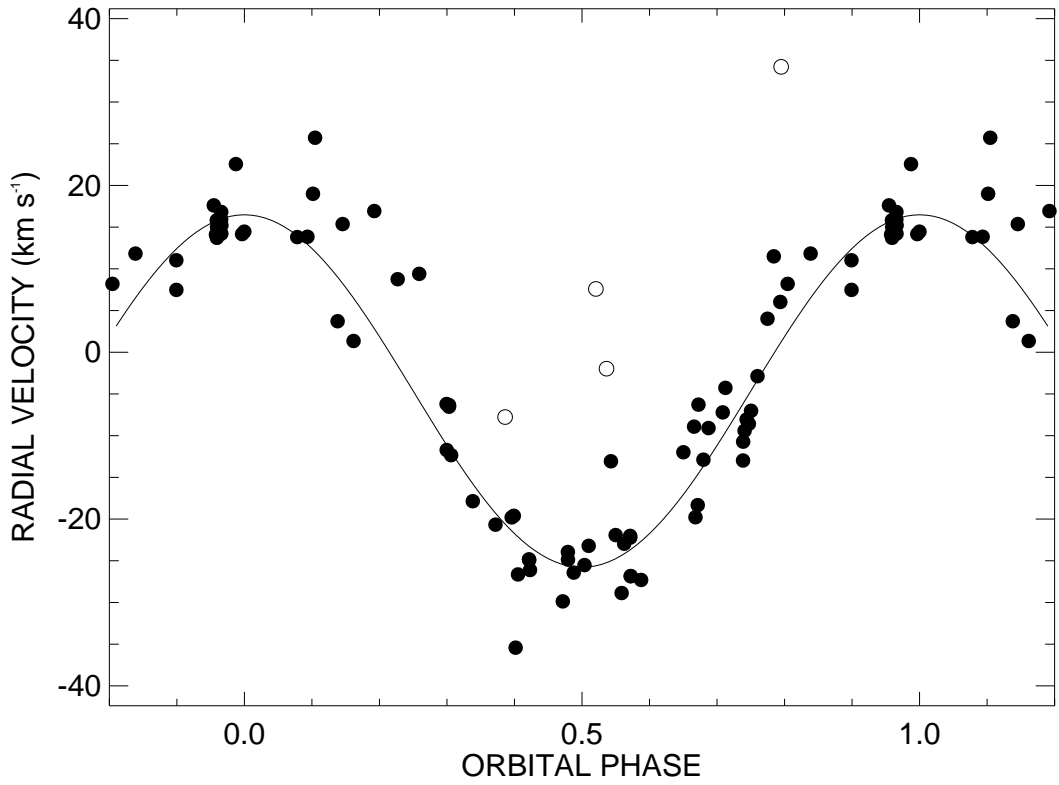}
\end{center}
\caption{The derived radial velocity curve of Ab1 (solid line) in the 312~d orbit. 
Open circles indicate those four measurements assigned zero 
weight in the solution.  Phase zero corresponds to the time
of maximum radial velocity (star crossing the ascending node) in 
this circular orbit.\label{fig7}}
\end{figure}

\placefigure{fig7}     % Figure 7 - radial velocity curve 312 d

\begin{figure}
\begin{center}
\includegraphics[angle=90,width=15cm]{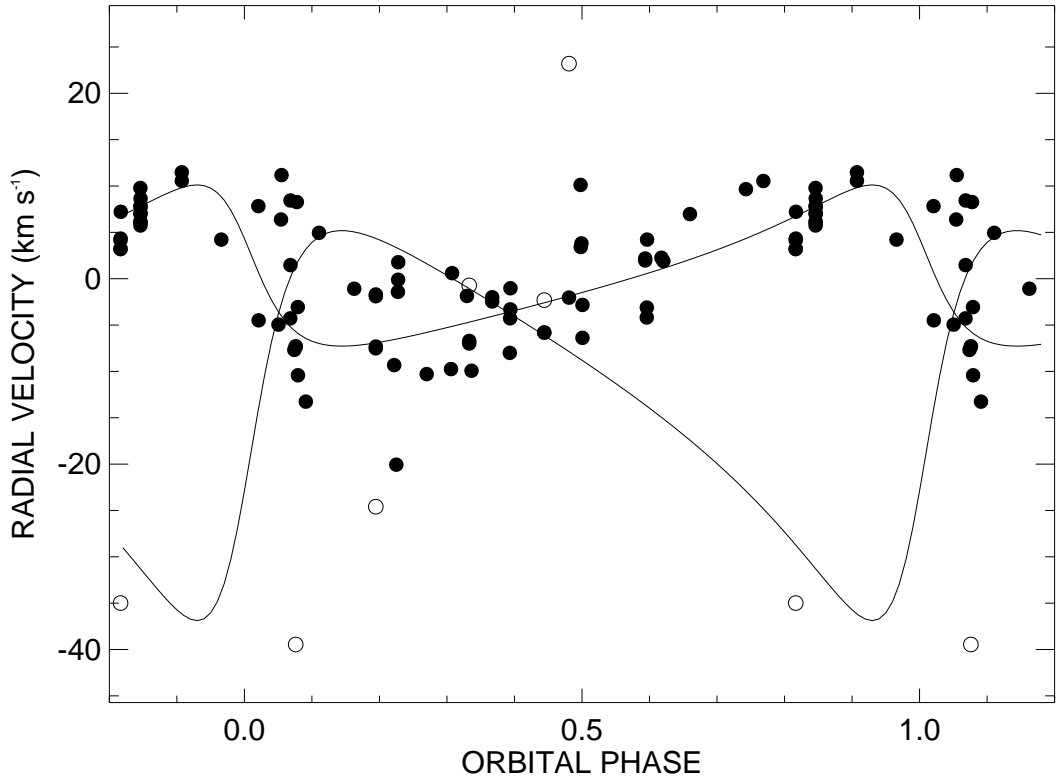}
\end{center}
\caption{The derived radial velocity curve of Ab1,Ab2 (low amplitude, solid line) 
in the 34~y orbit.  Open circles indicate measurements of the broad-lined Aa 
component and its preliminary radial velocity curve (large amplitude, solid
line).  Phase zero corresponds to the time of periastron.\label{fig8}}
\end{figure}

\placefigure{fig8}     % Figure 8 - radial velocity curve 34 y

\placetable{tab6}      % Table 6 - Ab1 orbital solutions

%%%%%%%%%%%%%%%%%%%%%%%%%%%%%%%%%%%%%%%%%%%%%%%%%%%%%%%%%%%%%%%

\section{Spectroscopic Properties}                  % Section 5

The two brightest components of HD~193322, Aa and Ab1, have very
broad and very narrow spectral lines, respectively, and indeed 
it is these properties that can help us distinguish their
different orbital motions.  We show in Figure~9 CFHT spectra 
of the \ion{He}{1} $\lambda 5876$ profile from 1986 and 2008. 
During this interval, the broad component moved slightly 
redward as expected for the anti-phase velocity curve of 
Aa between wide orbit phases 0.82 and 0.44 (Fig.~8).  
We collected all the available red spectra that recorded 
\ion{He}{1} $\lambda 5876$, and we formed an average spectrum 
for each run in order to increase the S/N of the spectra at 
each epoch.  We then formed model spectra for each of Aa and 
Ab1 from the grid of Lanz \& Hubeny (2003) using projected 
rotational velocities and model parameters optimized 
to match the composite profile (see below).  These model 
profiles were fit to the observations using a non-linear, 
least-squares procedure to derive the radial velocities of 
Aa and Ab1 at these epochs.  The derived Ab1 velocities are
identical within errors to the corresponding measurements 
given in Table~5, and the velocities for Aa are listed in 
Table~7.  The errors associated with the velocities of Aa 
are large, $\pm 20$ km~s$^{-1}$, because this component is 
so broad and shallow and because the shape of the red wing is
sensitive to the details of the removal of the telluric 
features found there.   Given these larger errors and the 
relatively small number of measurements, we made a constrained 
fit of the orbital radial velocity curve of Aa by setting 
all the parameters from the solution for Ab (Table~6) with 
the exception of the systemic velocity $\gamma$ and semiamplitude $K$, 
and by assigning a weight to each observation proportional to
the product of the spectral resolving power and 
the net S/N ratio in the adjoining continuum (column 5 of Table~7). 
The fit (illustrated in Fig.~8) yielded
$\gamma = -12 \pm 8$ km~s$^{-1}$ and $K = 21 \pm 16$ km~s$^{-1}$, 
with a residual rms $= 21$ km~s$^{-1}$.  These measurements 
are consistent with the expected Doppler shifts and masses for the 
Aa,Ab system (\S6).  

\begin{figure}
\begin{center}
\includegraphics[angle=90,width=15cm]{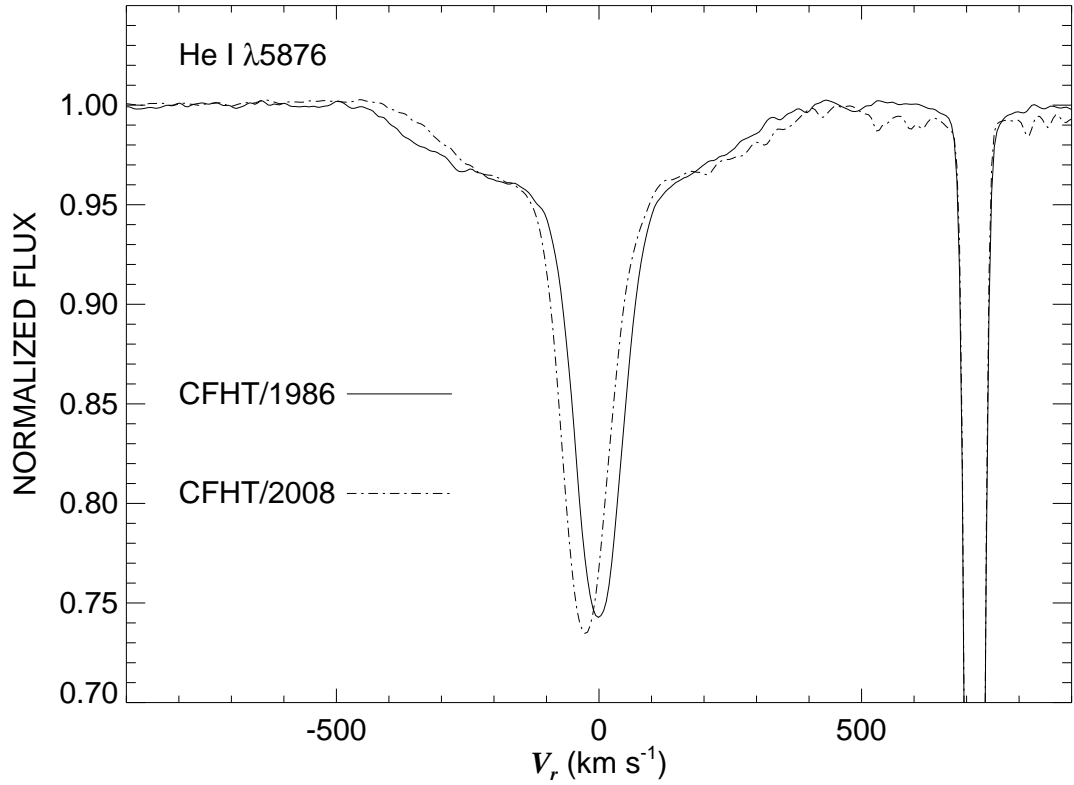}
\end{center}
\caption{CFHT spectroscopy of the \ion{He}{1} $\lambda 5876$ line profile 
from two epochs.  The narrow-lined component is associated with Ab1, while
the broad-lined component corresponds to Aa.  The interstellar \ion{Na}{1}
$\lambda 5890$ D2 line appears near $V_r = +730$ km~s$^{-1}$. 
\label{fig9}}
\end{figure}

\placefigure{fig9}     % Figure 9 - He I 5876 samples 

\placetable{tab7}      % Table 7 - Aa velocities

We estimated the spectroscopic parameters for the components 
by first reconstructing their spectra using the orbital velocity 
curves and a Doppler tomography algorithm (Bagnuolo et al.\ 1994) 
and then comparing the reconstructions with models from the 
grids of Lanz \& Hubeny (2003, 2007).  We selected nine spectra from 
our observations that recorded the blue portion of the spectrum 
with a resolving power greater than 10000 and that covered the 
extremes of motion in the wide and close systems (samples 10, 12, 
14, 15, and 16 from Table~4).  We began by running the tomography 
algorithm for only two components, Aa and Ab1, however, we found that 
the subsequent reconstructed spectrum for Aa had a composite appearance with 
both broad and narrow components, unlike our expectation from the
\ion{He}{1} $\lambda 5876$ profiles (Fig.~9).  We think that this 
is due to flux contamination in our blue spectra from the nearby 
B component (B1.5~V; $V\sin i \approx 100$ km~s$^{-1}$; 
see McKibben et al.\ 1998, Fig.~2, and Roberts et al.\ 2010, Fig.~1).  
Although component B may be a spectroscopic binary with a low semiamplitude  
(McKibben et al.\ 1998), we simply assumed that it was stationary
and contributed $11\%$ of the total flux (Roberts et al.\ 2010)
in the next iteration of tomographic reconstruction.  The power of the 
tomography algorithm to derive reliable and high quality reconstructed 
spectra increases with the number of spectra and with the orbital velocity 
range and flux contribution of the components.  Unfortunately, 
in the case of our blue spectra of HD~193322, these criteria are really 
only met for component Ab1.  The velocity range of Aa, for example, 
is so small relative to its characteristic line width that the 
algorithm may incorrectly assign line flux between the reconstruction 
of Aa and the stationary component B.  We dealt with this problem by
starting the initial guess for components Aa, Ab2, and B with 
model spectra rather than assuming a flat continuum spectrum (as done for Ab1). 
Although the resulting solutions are guided by our assumptions, 
they do at least show that the observed spectra are consistent with 
these assumptions since otherwise the reconstructed spectra would 
converge to an appearance different from the initial model guesses. 

We show the results of the full, four-component, tomographic reconstructions
in Figure~10.   These representative solutions were made using the orbital 
solutions from Table~6, adopting a mass ratio of 
$M({\rm Ab2}) / M({\rm Ab1}) = 0.37$ (\S6) 
and the flux ratios given in Table~8.  These flux ratios were calculated
from the $K^\prime$-band $\triangle m_{\rm close}$ and
$\triangle m_{\rm wide}$ results (\S3) assuming that these hot stars
all contribute by the same proportions in the $B$-band.  The spectroscopic
parameters were determined by finding the Lanz \& Hubeny (2003, 2007) model 
that best matched the absorption line ratios and H$\gamma$ Stark broadening 
(dependent on $T_{\rm eff}$ and $\log g$) and with a $V\sin i$ value adjusted 
to fit the widths of the absorption lines other than H$\gamma$.  The 
results are listed in columns 2 and 3 of Table~8 for Aa and Ab1, respectively, 
and we estimate that the associated errors are $\triangle T_{\rm eff} = \pm 1$~kK,
$\triangle \log g = \pm 0.5$, and $\triangle V\sin i = \pm 40$ and $\pm 10$ 
km~s$^{-1}$ for Aa and Ab1, respectively.   These parameters suggest spectral
classifications of O9~Vnn and O8.5~III for Aa and Ab1, respectively, based upon 
the calibration of Martins et al.\ (2005).  The ``nn'' suffix for the former 
classification indicates very broad lines.  
The relatively good agreement between the observed and model line depths
indicates that the flux ratios from interferometry (\S3) are fully 
consistent with the derived strengths of the spectroscopic features.  
The parameters in Table~8 for component B were taken from the work of 
Roberts et al.\ (2010), and the predicted model spectrum agrees well 
with the narrow, stationary spectral component from the tomographic 
reconstruction.  The results for the faintest component, Ab2, are poorly 
constrained because this star contributes such a small fraction of 
the total flux, but its spectrum suggests an early-B, dwarf classification.
We used the flux ratio between Ab2 and B and the temperature of B and
the theoretical main sequence $(T_{\rm eff}, M_V)$ adopted by 
Roberts et al.\ (2010) in order to estimate the effective temperature
of Ab2, assuming it is a main sequence star.  The H$\gamma$ $\lambda 4340$
line is the only strong feature in the reconstructed spectrum of Ab2, 
and the relative weakness of the \ion{He}{1} $\lambda\lambda 4387, 4471$
lines suggest that Ab2 may also be a rapid rotator with broad and 
shallow lines.  Note that the $V\sin i$ estimate for Ab2 in Table~8 is 
only approximate and may be subject to significant revision. 

\begin{figure}
\begin{center}
\includegraphics[angle=90,width=16cm]{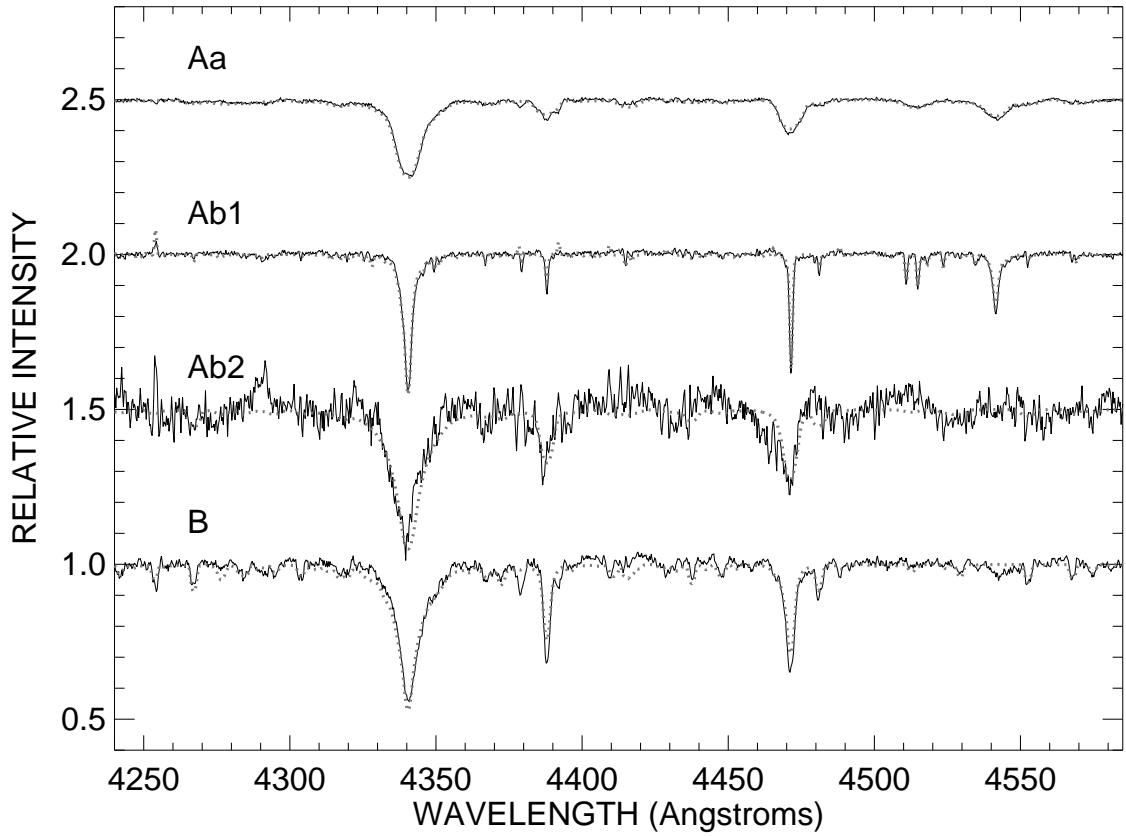}
\end{center}
\caption{Example tomographic reconstructions of the blue spectra 
of the components of HD~193322.  The solid lines show the 
Doppler tomography spectra while the dotted gray lines show 
superimposed model spectra.  The spectra are offset by 
steps of $50\%$ of the continuum for clarity.
\label{fig10}}
\end{figure}

\placefigure{fig10}    % Figure 10 - blue tomography results

\placetable{tab8}      % Table 8 - stellar parameters

%%%%%%%%%%%%%%%%%%%%%%%%%%%%%%%%%%%%%%%%%%%%%%%%%%%%%%%%%%%%%%%

\section{Discussion}                                % Section 6

One of the primary goals of this study was to determine the masses
of the component stars.  Since most of our results are preliminary, 
we cannot yet derive accurate masses, but the observational work 
does demonstrate the potential for improvement with further 
interferometric and spectroscopic observations.  
The mass sums for the wide and 
close systems can be determined from the angular semimajor axes 
and orbital periods (Table 2) and the distance $d=741 \pm 36$ pc
for Collinder~419 (Roberts et al.\ 2010).  The mass sums (eq.~2) are 
$(M({\rm Aa})+M({\rm Ab1})+M({\rm Ab2}))/M_\odot = 
(53.2 \pm 11.5) ~(d/[741~{\rm pc}])^3$ for the wide system and 
$(M({\rm Ab1})+M({\rm Ab2}))/M_\odot = 
35.6 ~(a/[4.0~{\rm mas}])^3 ~(d/[741~{\rm pc}])^3$ 
for the close system (where $a$ is the angular semimajor axis of Ab1,Ab2).  

To obtain the individual masses, we need to explore the 
solution space from the interferometric observations of 
the close binary (Fig.~3) and from the spectroscopic
orbit of Ab1 (Table~6).  In particular, we can use the location of 
the $\chi^2$ valley in the $(i,a)$ plane of Figure~3 to derive 
a family of solutions based solely upon $a$ (since $i = i(a)$ from Fig.~3).  
We show the derived individual masses as a function of 
$a$ in Figure~11.  The mass of Aa is set from the difference
of the total mass of Aa,Ab and the mass of Ab1,Ab2 (from $a$ and eq.~2);  
the mass of Ab2 is from eq.~3 (from $a$, $K$(Ab1), and the $i=i(a)$ relation in Fig.~3); 
and the mass of Ab1 is from the difference $M({\rm Ab1,Ab2}) - M({\rm Ab2})$.
These are all plotted in Figure~11 surrounded by a gray zone 
corresponding to the acceptable range in cluster distance.
We see that there is a strict upper 
limit of $a< 4.57$~mas required to keep $M({\rm Aa})>0$.
Furthermore, we also see that while the masses of Aa and Ab1
cover a significant range, the mass of Ab2 changes little 
over the range in $a$.  This is also shown in Figure~3, 
where the location of the $\chi^2$ valley is close to a 
contour of constant $M({\rm Ab2})$.  

\begin{figure}
\begin{center}
\includegraphics[angle=90,width=16cm]{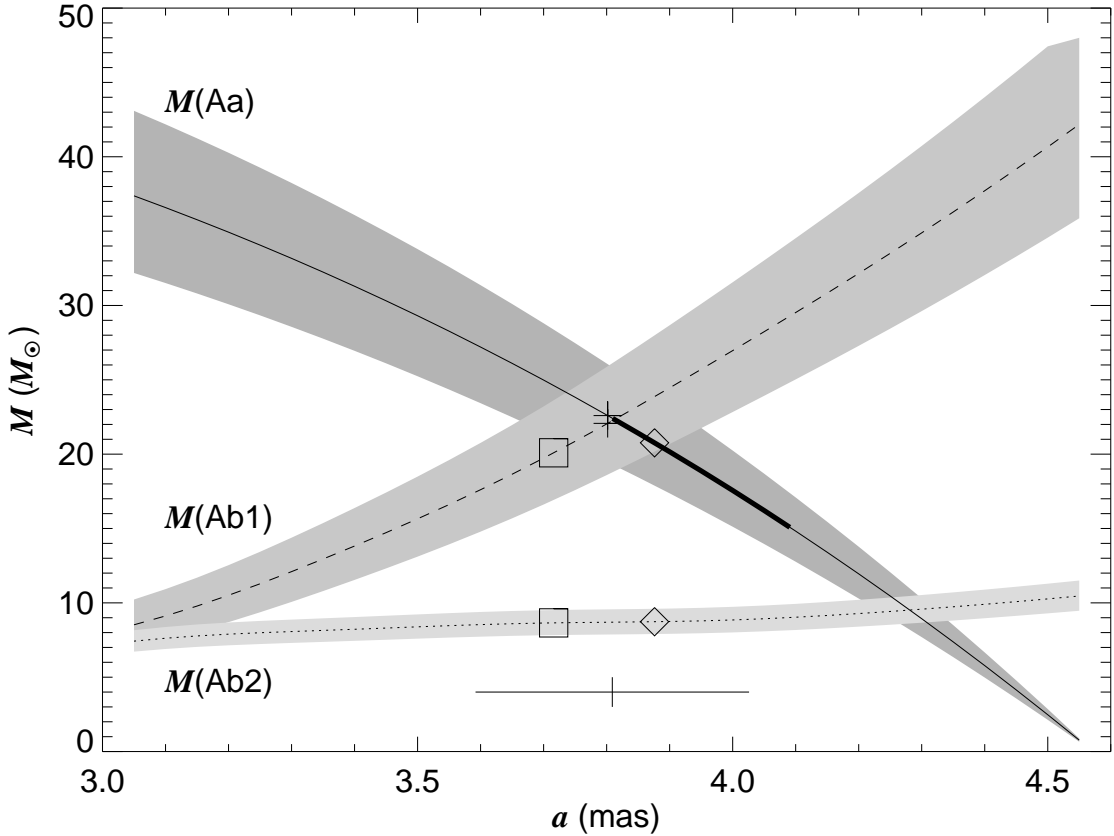}
\end{center}
\caption{A diagram of the mass solutions for components Aa (solid line), 
Ab1 (dashed line), and Ab2  (dotted line) as a function of $a$, the angular 
semimajor axis of the Ab1,Ab2 system.  The shaded region surrounding each 
line corresponds to the $\pm 1 \sigma$ error range in distance.
The thick portion of the Aa line shows the section that 
intersects with the $\pm 1 \sigma$ error range for the mass of Aa 
as determined from the visual wide orbit, assumed distance, 
and the Ab1,Ab2 center-of-mass radial velocity curve (Fig.~8).  
The various symbols indicate the positions where the mass ratios match 
those of main sequence stars with the observed flux ratios 
(squares for $F$(Ab2)/$F$(Ab1), crosses for $F$(Ab1)/$F$(Aa), and 
diamonds for $F$(Ab2)/$F$(Aa)).  The tick mark at bottom indicates 
those masses for which the sum of the corresponding fluxes of main sequence 
stars attains a minimum, the situation most consistent with the estimated
total absolute magnitude, and the horizontal line segment shows the range
over which the absolute magnitude is within 0.1 mag of the faint limit. 
\label{fig11}}
\end{figure}

\placefigure{fig11}    % Figure 11 - (a, mass) diagram

Another constraint on the mass of Aa can be formed independent of 
the details of the Ab1,Ab2 orbit by applying eq.~3 to the 
wide orbit.  We take $a$, $P$, $e$, and $i$ from the visual 
orbit of the wide system (Table~2) and combine these with the
orbital semiamplitude $K({\rm Ab})$ from spectroscopy (Table~6)
to obtain $M({\rm Aa}) / M_\odot = (18.7 \pm 3.7) (d/[741~{\rm pc}])^2$.  
The $\pm 1 \sigma$ region from this relation is plotted as 
the thick line segment for $M({\rm Aa})$ in Figure~11, and 
it corresponds to a range in semimajor axis of $a = 3.96 \pm 0.14$~mas.

It is reasonable to assume that all three stars 
are main sequence objects given the position of the Aa,Ab 
system in the color-magnitude diagram (Roberts et al.\ 2010). 
The $K$-band fluxes of massive main sequence stars scale with mass
$M$ as $F \propto M^{2.30}$ for stars in this mass range 
according to the models of Marigo et al.\ (2008), 
so we can use this relation to predict the flux ratio 
between any pair of stars according to the mass relations
shown in Figure~11.  The positions where the model flux 
ratios match the observed ones (Table~8) are indicated 
by pairs of symbols in Figure~11.  These flux ratio 
relations indicate a semimajor axis range of $a = 3.80 \pm 0.08$~mas.

A final constraint can be set from the overall fluxes 
and absolute magnitude of the combined system.   
Ten Brummelaar et al.\ (2000) estimate that the apparent 
$V$-band magnitude of Aa,Ab is $V = 5.96 \pm 0.02$ mag, 
and using the distance and extinction for the star from 
Roberts et al.\ (2010), we estimate that the absolute 
magnitude is $M_V = -4.35 \pm 0.12$ mag.  The individual 
absolute magnitudes of the components are given in Table~8.  
We can apply the mass -- absolute magnitude relation $(M,M_V)$ for the main sequence from 
the models of Marigo et al.\ (2008) to obtain a prediction for the 
total absolute magnitude for each of the mass combinations 
shown in Figure~11.  The best match occurs for the masses 
obtained at $a=3.81$~mas, approximately where Aa and Ab1 
have equal masses.  For those masses the predicted absolute magnitude 
is $M_V = -5.12$, which is significantly brighter 
than the estimate above from observations.  The models
predict even brighter fluxes if either of Aa or Ab1 are more massive, 
and the horizontal line at the bottom of Figure~11 shows 
the range in $a$ where the combined magnitude is within 0.1~mag 
of the faint limit.  

The average estimate for the semimajor axis from the above 
three constraints is $a=3.85 \pm 0.09$~mas, and we adopt 
the associated mass solution as best representing 
the current observational data.  
The masses and other properties summarized in Table~8 are 
generally in agreement with expectations for hot, main sequence
stars (Martins et al.\ 2005).  However, there remain a number
of significant discrepancies that deserve further investigation. 
The mass of component Ab1 is similar to that expected for an 
O8.5~III star ($\approx 24 M_\odot$; Martins et al.\ 2005), but  
the star's absolute magnitude is about 1.8 mag fainter 
than typical for such stars.  This discrepancy hints that 
Ab1 may be a dwarf rather than a giant star.  
The overall faintness of the system compared with 
expectations for the stars' masses may indicate that
the distance estimate needs to be revised downward (leading to 
lower masses) and/or the extinction estimate revised upwards. 
There also remains some confusion about which of Aa or Ab1,Ab2 
is brighter.  As we noted in \S3, the visibility analysis 
indicates that $\triangle m_{\rm wide} = m({\rm Aa}) - m({\rm Ab})
= 0.086 \pm 0.012$ mag (Ab brighter than Aa), which agrees 
within errors with high angular resolution measurements 
using the AstraLux camera by Ma\'{i}z Apell\'{a}niz (2010),  
$m({\rm Aa}) - m({\rm Ab}) = 0.04 \pm 0.19$ mag. 
Although these results indicate that Ab is somewhat brighter than Aa, 
we refrain from re-designating the identities of the components
to avoid confusion with published results.  

We expect that these lingering problems will be resolved with 
future interferometric observations that will better sample the 
orbital phases of the close pair near the nodal crossings and 
will lead to improved constraints on the angular semimajor axis $a$. 
In addition, we clearly need to continue the long term high resolution 
work on the wide orbit to cover the missing orbital phases (Fig.~2).
We plan to obtain these measurements through continuing observations
of this system using the CHARA Array interferometer. 
Additional high S/N and high resolution spectroscopy holds the 
promise to deliver better orbital constraints on Aa and Ab2 
that would then allow us to estimate the masses without 
relying on the distance of the cluster.  Indeed, reliable 
orbital elements would render it possible to set an independent 
estimate of the cluster distance.  We anticipate expanding the 
spectroscopic coverage over the next decade.  

We close with some speculative remarks about the angular momentum
distribution of the stars of HD~193322.  It is remarkable 
that this system contains both a very rapidly rotating star (Aa) 
and a very slowly rotating star (Ab1).  It is possible that Ab1  
has a rotational axis with a low inclination, so that its equatorial 
rotational velocity is close to typical values.  However, 
the very large line broadening of Aa places it among the most rapidly 
rotating O-type stars known (Penny 1996).  It is possible that the 
angular momentum of the natal cloud led directly to rapid rotation 
in the case of Aa and to the formation of a binary in the case of Ab. 
Alternatively, there may have been some very close gravitational 
encounters in the early life of the system.  In some circumstances, 
a close encounter between a binary and a third interloper can lead
to a merger of two of the components and ejection of the third 
(Gaburov et al.\ 2010).  It is possible that the rapid rotator Aa 
is such a merger product and that the runaway star 68~Cygni 
was the object ejected from the system (Schilbach \& R\"{o}ser 2008). 
If so, then the orbital and spin properties of the stars of HD~193322
offer key evidence about the early dynamical processes in this cluster.

%%%%%%%%%%%%%%%%%%%%%%%%%%%%%%%%%%%%%%%%%%%%%%%%%%%%%%%%%%%%%%%

\acknowledgments

The CHARA Array, operated by Georgia State University, was built 
with funding provided by the National Science Foundation, 
Georgia State University, the W.\ M.\ Keck Foundation, and the 
David and Lucile Packard Foundation.  This material is based 
upon work supported by the National Science Foundation 
under Grants No.~AST-0606861, AST-0606958, AST-0908253, and AST-1009080. 
Institutional support has been provided from the GSU College
of Arts and Sciences and from the Research Program Enhancement
fund of the Board of Regents of the University System of Georgia,
administered through the GSU Office of the Vice President
for Research.  B.D.M.\ and W.I.H.\ have been supported by the 
National Aeronautics and Space Administration under reimbursable no.\ NNH06AD70I, 
issued through the Terrestrial Planet Finder Foundation Science program. 
Thanks are also extended to Ken Johnston and the U.\ S.\ Naval Observatory for 
their continued support of the Double Star Program. 
M.V.M.\ thanks Lehigh University for an institutional grant.
A portion of the research in this paper was carried out at the Jet
Propulsion Laboratory, California Institute of Technology,
under a contract with the National Aeronautics and Space Administration.  
We are grateful to Dr.\ Gregg Wade and the MiMeS consortium 
for sharing with us their spectral data on HD~193322 in advance
of publication.  This work is partially based on spectral data retrieved 
from the ELODIE archive at Observatoire de Haute-Provence (OHP). 
Additional spectroscopic data were retrieved from Ritter  
Observatory's public archive, which was supported by the National  
Science Foundation Program for Research and Education with Small  
Telescopes (NSF-PREST) under grant AST-0440784.

%%%%%%%%%%%%%%%%%%%%%%%%%%%%%%%%%%%%%%%%%%%%%%%%%%%%%%%%%%%%%%%

% Bibliography 

%%%%%%%%%%%%%%%%%%%%%%%%%%%%%%%%%%%%%%%%%%%%%%%%%%%%%%%%%%%%%%%

% Tables

%Table 1: Aa,Ab measurements 

\begin{deluxetable}{ccccc}
\tablewidth{0pc}
\tablenum{1}
\tablecaption{Astrometric Measurements of Aa,Ab\label{tab1}}
\tablehead{
\colhead{Date} &
\colhead{$\theta$} &
\colhead{$\rho$} &
\colhead{Data} &
\colhead{} \\
\colhead{(BY)} &
\colhead{(deg)} &
\colhead{(arcsec)} &
\colhead{Type} &
\colhead{Reference} 
}
\startdata
1985.5177                  &  188.4    &     0.049\phn &     Speckle  & McAlister et al.\ (1993) \\
1985.8396                  &  192.5    &     0.049\phn &     Speckle  & McAlister et al.\ (1993) \\ 
1986.8884                  &  198.6    &     0.049\phn &     Speckle  & McAlister et al.\ (1993) \\
1988.6630                  &  216.6    &     0.048\phn &     Speckle  & McAlister et al.\ (1993) \\
1989.7061                  &  229.6    &     0.045\phn &     Speckle  & McAlister et al.\ (1993) \\
2005.6054                  &  109.1    &     0.0638    &     OLBI/SFP & This paper               \\
2005.7350                  &  107.7    &     0.0647    &     OLBI/SFP & This paper               \\
2005.8652\tablenotemark{a} &  100.4    &     0.086\phn &     Speckle  & Mason et al.\ (2009)     \\
2006.4324\tablenotemark{a} &  100.1    &     0.0409    &     OLBI/SFP & This paper               \\
2006.4897                  &  101.5    &     0.0670    &     OLBI/SFP & This paper               \\
2006.5881                  &  113.9    &     0.0651    &     OLBI/SFP & This paper               \\
2006.6758                  &  118.0    &     0.0565    &     OLBI/SFP & This paper               \\
2007.4729                  &  111.7    &     0.0666    &     OLBI/SFP & This paper               \\
2007.5098                  &  113.6    &     0.0665    &     OLBI/SFP & This paper               \\
2007.6042\tablenotemark{a} &  100.9    &     0.067\phn &     Speckle  & Mason et al.\ (2009)     \\
2008.4508                  &  116.8    &     0.066\phn &     Speckle  & Mason et al.\ (2009)     \\
2008.6198                  &  121.8    &     0.0616    &     OLBI/SFP & This paper               \\
2008.8028                  &  124.7    &     0.0551    &     OLBI/SFP & This paper               \\
2009.4178                  &  120.1    &     0.0626    &     OLBI/SFP & This paper               \\
2009.5017                  &  126.7    &     0.0575    &     OLBI/SFP & This paper               \\
2009.6146                  &  122.0    &     0.0651    &     OLBI/SFP & This paper               \\
2009.7776                  &  122.0    &     0.0649    &     OLBI/SFP & This paper               \\
2010.8753                  &  129.8    &     0.0648    &     OLBI/SFP & This paper               
\enddata
\tablenotetext{a}{Assigned zero weight in the fit.}
\end{deluxetable}

%Table 2: Angular Orbital Elements
\begin{deluxetable}{lcc}
\tablewidth{0pc}
\tablenum{2}
\tablecaption{Visual Orbital Elements\label{tab2}}
\tablehead{
\colhead{Element} &
\colhead{Aa,Ab Orbit} &
\colhead{Ab1,Ab2 Orbit}}
\startdata
$P$ (y)               \dotfill & $35.20 \pm 1.45$   &  0.85533\tablenotemark{a}  \\
$P$ (d)               \dotfill & $12855 \pm 528$    &  312.40\tablenotemark{a}   \\
$T$ (BY)              \dotfill & $1994.84 \pm 1.69$ &  1996.109\tablenotemark{a} \\
$T$ (HJD--2,400,000)  \dotfill & $49662 \pm 616$    &  50123.5\tablenotemark{a}  \\
$a$ (mas)             \dotfill & $54.5 \pm 3.7$     &  $3.9^{+1.1}_{-1.2}$       \\
$i$ (deg)             \dotfill & $46.2 \pm 6.9$     &  $51^{+17}_{-51}$          \\
$\Omega$ (deg)        \dotfill & $255.2 \pm 15.0$   &  $25^{+3}_{-35}$\tablenotemark{b} \\
$e$                   \dotfill & $0.489 \pm 0.081$  &  0\tablenotemark{a}        \\
$\omega$ (deg)        \dotfill & $70.4 \pm 7.5$     &  180\tablenotemark{c}   
\enddata
\tablenotetext{a}{Fixed with values from the radial velocity orbit (Table 6).}
\tablenotetext{b}{Or $205^{+3}_{-35}$ deg.}
\tablenotetext{c}{Fixed for the relative orbit of Ab2 with respect to Ab1.}
\end{deluxetable}

%Table 3: SFP visibility measurements

\clearpage 
\begin{deluxetable}{ccccccc} 
\tablewidth{0pc} 
\tabletypesize{\small} 
%\rotate 
\tablenum{3} 
\tablecaption{Visibility Measurements of Ab1,Ab2 \label{tab3}} 
\tablehead{ 
\colhead{Date} & 
\colhead{$\phi$} & 
\colhead{$B_p$} & 
\colhead{$\psi_p$} & 
\colhead{} & 
\colhead{} & 
\colhead{} \\ 
\colhead{(HJD$-$2,400,000)} & 
\colhead{(close)} & 
\colhead{(m)} & 
\colhead{(deg)} & 
\colhead{$V$} & 
\colhead{$\triangle V$} & 
\colhead{$O-C$}} 
\startdata 
53591.721 & 0.102 & \phn98.8 &       118.8 &  1.035 &  0.175 & \phs0.045 \\
53591.756 & 0.102 &    104.3 &       109.0 &  0.968 &  0.147 &  $-$0.020 \\
53591.785 & 0.102 &    107.0 &       101.9 &  0.912 &  0.125 &  $-$0.024 \\
53591.818 & 0.102 &    107.9 & \phn   93.9 &  0.812 &  0.108 &  $-$0.039 \\
53591.842 & 0.102 &    106.9 & \phn   88.5 &  0.697 &  0.095 &  $-$0.104 \\
53591.855 & 0.102 &    105.7 & \phn   85.3 &  0.740 &  0.097 &  $-$0.038 \\
53591.888 & 0.102 &    101.1 & \phn   77.2 &  0.679 &  0.093 &  $-$0.073 \\
53591.912 & 0.102 & \phn96.2 & \phn   70.6 &  0.667 &  0.088 &  $-$0.084 \\
53638.741 & 0.252 &    170.6 &       324.4 &  0.893 &  0.242 & \phs0.137 \\
53638.747 & 0.252 &    169.8 &       323.5 &  0.702 &  0.119 &  $-$0.057 \\
53638.753 & 0.252 &    168.9 &       322.4 &  0.853 &  0.153 & \phs0.090 \\
53638.757 & 0.252 &    168.2 &       321.7 &  0.797 &  0.150 & \phs0.031 \\
53638.765 & 0.252 &    166.8 &       320.4 &  0.724 &  0.129 &  $-$0.047 \\
53638.769 & 0.252 &    166.1 &       319.8 &  0.834 &  0.160 & \phs0.061 \\
53638.775 & 0.252 &    164.9 &       318.9 &  0.802 &  0.170 & \phs0.025 \\
53638.779 & 0.252 &    164.0 &       318.3 &  0.778 &  0.173 &  $-$0.001 \\
53638.782 & 0.252 &    163.3 &       317.8 &  0.940 &  0.169 & \phs0.159 \\
53638.787 & 0.252 &    162.2 &       317.1 &  1.023 &  0.145 & \phs0.240 \\
53638.791 & 0.252 &    161.2 &       316.5 &  0.920 &  0.174 & \phs0.135 \\
53638.794 & 0.252 &    160.6 &       316.1 &  0.780 &  0.162 &  $-$0.007 \\
53638.798 & 0.252 &    159.4 &       315.4 &  0.884 &  0.166 & \phs0.096 \\
53638.804 & 0.252 &    158.0 &       314.7 &  0.817 &  0.174 & \phs0.027 \\
53638.809 & 0.252 &    156.5 &       314.0 &  0.938 &  0.155 & \phs0.148 \\
53638.813 & 0.253 &    155.4 &       313.5 &  0.931 &  0.218 & \phs0.140 \\
53639.696 & 0.255 &    107.6 & \phn   91.8 &  0.831 &  0.118 & \phs0.057 \\
53639.701 & 0.255 &    107.4 & \phn   90.6 &  0.778 &  0.119 & \phs0.005 \\
53639.705 & 0.255 &    107.2 & \phn   89.6 &  0.817 &  0.120 & \phs0.045 \\
53639.713 & 0.255 &    106.7 & \phn   87.8 &  0.901 &  0.140 & \phs0.130 \\
53639.717 & 0.255 &    106.3 & \phn   86.8 &  0.775 &  0.111 & \phs0.005 \\
53639.723 & 0.255 &    105.7 & \phn   85.3 &  0.765 &  0.118 &  $-$0.002 \\
53639.729 & 0.255 &    105.0 & \phn   83.8 &  0.594 &  0.087 &  $-$0.171 \\
53639.732 & 0.255 &    104.6 & \phn   83.1 &  0.722 &  0.122 &  $-$0.041 \\
53639.736 & 0.255 &    104.2 & \phn   82.2 &  0.732 &  0.106 &  $-$0.031 \\
53639.739 & 0.255 &    103.8 & \phn   81.5 &  0.829 &  0.159 & \phs0.068 \\
53639.743 & 0.255 &    103.2 & \phn   80.5 &  0.785 &  0.116 & \phs0.026 \\
53639.746 & 0.255 &    102.6 & \phn   79.6 &  0.888 &  0.173 & \phs0.130 \\
53639.750 & 0.256 &    102.0 & \phn   78.6 &  0.834 &  0.145 & \phs0.078 \\
53639.754 & 0.256 &    101.4 & \phn   77.6 &  0.638 &  0.097 &  $-$0.117 \\
53639.757 & 0.256 &    100.9 & \phn   76.9 &  0.734 &  0.110 &  $-$0.020 \\
53639.767 & 0.256 & \phn99.0 & \phn   74.3 &  0.792 &  0.148 & \phs0.040 \\
53639.769 & 0.256 & \phn98.5 & \phn   73.6 &  0.729 &  0.128 &  $-$0.022 \\
53639.774 & 0.256 & \phn97.5 & \phn   72.3 &  0.848 &  0.198 & \phs0.098 \\
53639.778 & 0.256 & \phn96.7 & \phn   71.3 &  0.611 &  0.116 &  $-$0.139 \\
53639.781 & 0.256 & \phn96.0 & \phn   70.3 &  0.666 &  0.113 &  $-$0.085 \\
53639.785 & 0.256 & \phn95.1 & \phn   69.2 &  0.832 &  0.170 & \phs0.081 \\
53639.788 & 0.256 & \phn94.4 & \phn   68.3 &  0.917 &  0.147 & \phs0.166 \\
53639.793 & 0.256 & \phn93.3 & \phn   66.9 &  0.938 &  0.198 & \phs0.185 \\
53639.797 & 0.256 & \phn92.3 & \phn   65.7 &  1.270 &  0.244 & \phs0.515 \\
53639.800 & 0.256 & \phn91.7 & \phn   64.9 &  1.379 &  0.290 & \phs0.622 \\
53639.808 & 0.256 & \phn89.7 & \phn   62.4 &  1.003 &  0.240 & \phs0.240 \\
53639.812 & 0.256 & \phn88.6 & \phn   61.1 &  0.776 &  0.159 & \phs0.008 \\
53639.816 & 0.256 & \phn87.4 & \phn   59.5 &  0.922 &  0.221 & \phs0.148 \\
53639.822 & 0.256 & \phn86.0 & \phn   57.7 &  1.040 &  0.222 & \phs0.258 \\
53893.877 & 0.069 &    315.5 & \phn   39.3 &  1.154 &  0.163 & \phs0.357 \\
53893.886 & 0.069 &    318.1 & \phn   37.7 &  0.989 &  0.144 & \phs0.209 \\
53893.889 & 0.069 &    319.0 & \phn   37.1 &  0.893 &  0.127 & \phs0.117 \\
53893.898 & 0.069 &    321.1 & \phn   35.5 &  0.839 &  0.130 & \phs0.068 \\
53893.912 & 0.069 &    323.8 & \phn   33.0 &  1.062 &  0.146 & \phs0.287 \\
53893.915 & 0.069 &    324.3 & \phn   32.4 &  1.088 &  0.154 & \phs0.309 \\
53893.923 & 0.069 &    325.6 & \phn   30.8 &  1.180 &  0.160 & \phs0.392 \\
53893.926 & 0.069 &    325.9 & \phn   30.3 &  1.031 &  0.136 & \phs0.240 \\
53893.935 & 0.069 &    327.0 & \phn   28.6 &  0.931 &  0.130 & \phs0.128 \\
53914.806 & 0.136 & \phn93.0 &       128.0 &  1.021 &  0.134 & \phs0.024 \\
53914.815 & 0.136 & \phn94.8 &       125.1 &  1.044 &  0.160 & \phs0.045 \\
53914.824 & 0.136 & \phn96.7 &       122.1 &  1.072 &  0.141 & \phs0.075 \\
53914.829 & 0.136 & \phn97.6 &       120.7 &  1.029 &  0.137 & \phs0.036 \\
53914.837 & 0.136 & \phn99.2 &       118.1 &  1.044 &  0.144 & \phs0.059 \\
53914.842 & 0.136 & \phn99.9 &       117.0 &  0.940 &  0.125 &  $-$0.038 \\
53914.850 & 0.136 &    101.4 &       114.5 &  0.925 &  0.125 &  $-$0.039 \\
53914.859 & 0.136 &    102.7 &       112.2 &  0.937 &  0.123 &  $-$0.009 \\
53914.865 & 0.136 &    103.6 &       110.4 &  0.872 &  0.124 &  $-$0.059 \\
53950.685 & 0.251 & \phn88.2 &       136.2 &  0.896 &  0.123 & \phs0.077 \\
53950.695 & 0.251 & \phn90.2 &       132.7 &  0.704 &  0.093 &  $-$0.095 \\
53950.701 & 0.251 & \phn91.5 &       130.5 &  0.773 &  0.103 &  $-$0.015 \\
53950.708 & 0.251 & \phn92.9 &       128.3 &  0.770 &  0.102 &  $-$0.007 \\
53950.716 & 0.251 & \phn94.5 &       125.7 &  0.692 &  0.090 &  $-$0.075 \\
53950.724 & 0.251 & \phn96.1 &       123.1 &  0.759 &  0.100 &  $-$0.000 \\
53950.730 & 0.251 & \phn97.3 &       121.1 &  0.687 &  0.091 &  $-$0.067 \\
53950.736 & 0.251 & \phn98.5 &       119.3 &  0.667 &  0.089 &  $-$0.085 \\
53950.744 & 0.251 &    100.0 &       116.9 &  0.733 &  0.096 &  $-$0.017 \\
53950.755 & 0.251 &    101.7 &       113.9 &  0.760 &  0.101 & \phs0.010 \\
53950.764 & 0.251 &    103.0 &       111.6 &  0.678 &  0.090 &  $-$0.074 \\
53950.781 & 0.251 &    105.2 &       107.1 &  0.797 &  0.105 & \phs0.039 \\
53950.788 & 0.251 &    105.9 &       105.3 &  0.778 &  0.104 & \phs0.017 \\
53950.803 & 0.251 &    107.1 &       101.7 &  0.849 &  0.122 & \phs0.082 \\
53950.814 & 0.251 &    107.7 & \phn   98.9 &  0.826 &  0.121 & \phs0.055 \\
53950.819 & 0.251 &    107.8 & \phn   97.9 &  0.729 &  0.105 &  $-$0.043 \\
53950.830 & 0.251 &    107.9 & \phn   95.3 &  0.719 &  0.103 &  $-$0.054 \\
53950.838 & 0.251 &    107.8 & \phn   93.2 &  0.774 &  0.103 &  $-$0.000 \\
53950.843 & 0.251 &    107.7 & \phn   92.2 &  0.829 &  0.114 & \phs0.055 \\
53950.852 & 0.251 &    107.3 & \phn   90.1 &  0.808 &  0.119 & \phs0.035 \\
53950.861 & 0.251 &    106.7 & \phn   88.0 &  0.720 &  0.096 &  $-$0.051 \\
53950.866 & 0.251 &    106.2 & \phn   86.6 &  0.839 &  0.114 & \phs0.069 \\
53950.875 & 0.251 &    105.3 & \phn   84.5 &  0.785 &  0.110 & \phs0.018 \\
53950.888 & 0.251 &    103.8 & \phn   81.5 &  0.751 &  0.104 &  $-$0.012 \\
53950.900 & 0.251 &    101.9 & \phn   78.4 &  0.836 &  0.112 & \phs0.079 \\
53950.904 & 0.251 &    101.3 & \phn   77.5 &  0.708 &  0.096 &  $-$0.048 \\
53982.762 & 0.353 &    174.5 &       331.3 &  1.133 &  0.154 & \phs0.199 \\
53982.769 & 0.353 &    174.0 &       330.0 &  1.068 &  0.147 & \phs0.135 \\
53982.777 & 0.354 &    173.2 &       328.4 &  1.070 &  0.142 & \phs0.137 \\
53982.782 & 0.354 &    172.7 &       327.4 &  1.088 &  0.141 & \phs0.156 \\
53982.791 & 0.354 &    171.8 &       326.0 &  1.017 &  0.136 & \phs0.083 \\
53982.797 & 0.354 &    171.0 &       324.9 &  1.017 &  0.135 & \phs0.081 \\
53982.805 & 0.354 &    169.8 &       323.5 &  1.008 &  0.135 & \phs0.069 \\
53982.814 & 0.354 &    168.5 &       322.1 &  0.975 &  0.135 & \phs0.031 \\
53982.818 & 0.354 &    167.8 &       321.3 &  1.018 &  0.136 & \phs0.072 \\
53982.827 & 0.354 &    166.1 &       319.9 &  1.053 &  0.138 & \phs0.100 \\
53982.849 & 0.354 &    161.4 &       316.6 &  0.677 &  0.110 &  $-$0.292 \\
53982.854 & 0.354 &    160.2 &       315.9 &  1.053 &  0.138 & \phs0.080 \\
53982.863 & 0.354 &    157.7 &       314.6 &  1.040 &  0.139 & \phs0.060 \\
54273.899 & 0.285 &    105.6 &       106.2 &  0.747 &  0.112 &  $-$0.032 \\
54273.910 & 0.285 &    106.6 &       103.5 &  0.702 &  0.103 &  $-$0.079 \\
54273.914 & 0.285 &    106.9 &       102.3 &  0.795 &  0.110 & \phs0.014 \\
54273.930 & 0.286 &    107.7 & \phn   98.5 &  0.798 &  0.106 & \phs0.018 \\
54273.938 & 0.286 &    107.9 & \phn   96.6 &  0.789 &  0.104 & \phs0.011 \\
54273.948 & 0.286 &    107.9 & \phn   94.4 &  0.789 &  0.105 & \phs0.013 \\
54273.958 & 0.286 &    107.7 & \phn   91.9 &  0.879 &  0.114 & \phs0.107 \\
54273.963 & 0.286 &    107.5 & \phn   90.8 &  0.794 &  0.103 & \phs0.024 \\
54273.979 & 0.286 &    106.3 & \phn   86.9 &  0.847 &  0.109 & \phs0.085 \\
54273.987 & 0.286 &    105.6 & \phn   85.1 &  0.785 &  0.101 & \phs0.027 \\
54273.997 & 0.286 &    104.4 & \phn   82.7 &  0.719 &  0.093 &  $-$0.035 \\
54285.938 & 0.324 &    247.9 & \phn\phn8.0 &  0.761 &  0.169 &  $-$0.130 \\
54288.939 & 0.334 &    247.9 & \phn\phn6.0 &  1.072 &  0.164 & \phs0.090 \\
54288.986 & 0.334 &    248.0 &       355.4 &  0.860 &  0.141 &  $-$0.080 \\
54289.971 & 0.337 &    248.0 &       358.2 &  0.755 &  0.115 &  $-$0.195 \\
54289.975 & 0.337 &    248.0 &       357.3 &  0.700 &  0.272 &  $-$0.240 \\
54318.890 & 0.429 &    330.7 & \phn\phn2.5 &  1.094 &  0.144 & \phs0.316 \\
54412.690 & 0.730 & \phn89.5 & \phn   62.2 &  1.096 &  0.226 & \phs0.345 \\
54412.715 & 0.730 & \phn82.9 & \phn   53.5 &  0.839 &  0.141 & \phs0.073 \\
54412.727 & 0.730 & \phn79.9 & \phn   49.1 &  0.716 &  0.115 &  $-$0.065 \\
54412.739 & 0.730 & \phn76.8 & \phn   44.2 &  0.801 &  0.131 &  $-$0.000 \\
54606.005 & 0.348 &    278.4 &       143.2 &  0.732 &  0.191 &  $-$0.218 \\
54657.944 & 0.515 &    267.3 &       127.0 &  0.731 &  0.156 &  $-$0.092 \\
54657.959 & 0.515 &    262.1 &       124.1 &  0.986 &  0.191 & \phs0.074 \\
54657.968 & 0.515 &    258.6 &       122.5 &  1.115 &  0.181 & \phs0.161 \\
54692.830 & 0.626 &    272.2 &       130.7 &  0.684 &  0.096 &  $-$0.147 \\
54692.837 & 0.626 &    270.6 &       129.3 &  0.727 &  0.096 &  $-$0.143 \\
54692.889 & 0.627 &    330.7 &       177.4 &  1.008 &  0.146 & \phs0.033 \\
54692.897 & 0.627 &    330.7 &       175.6 &  1.068 &  0.144 & \phs0.099 \\
54692.905 & 0.627 &    330.7 &       173.9 &  0.761 &  0.112 &  $-$0.196 \\
54692.912 & 0.627 &    330.7 &       172.0 &  0.702 &  0.127 &  $-$0.232 \\
54692.946 & 0.627 &    330.4 &       164.3 &  0.702 &  0.196 &  $-$0.087 \\
54692.960 & 0.627 &    330.1 &       161.2 &  0.580 &  0.128 &  $-$0.174 \\
54759.629 & 0.840 &    275.2 &       134.0 &  0.983 &  0.168 & \phs0.063 \\
54759.667 & 0.840 &    330.7 &       186.2 &  0.997 &  0.175 & \phs0.193 \\
54759.677 & 0.840 &    330.7 &       183.9 &  1.271 &  0.183 & \phs0.503 \\
54759.687 & 0.840 &    330.7 &       181.6 &  1.092 &  0.203 & \phs0.340 \\
54759.696 & 0.840 &    330.7 &       179.5 &  1.380 &  0.215 & \phs0.623 \\
54759.728 & 0.841 &    238.0 &       115.8 &  1.088 &  0.206 & \phs0.332 \\
54759.765 & 0.841 &    330.3 &       163.6 &  0.809 &  0.133 &  $-$0.151 \\
54759.790 & 0.841 &    329.5 &       158.0 &  0.816 &  0.133 &  $-$0.160 \\
54983.996 & 0.558 &    277.2 &       137.8 &  0.640 &  0.091 &  $-$0.172 \\
54984.002 & 0.558 &    276.7 &       136.6 &  0.766 &  0.100 &  $-$0.076 \\
54984.865 & 0.561 &    262.4 &       100.2 &  1.204 &  0.160 & \phs0.239 \\
54984.871 & 0.561 &    267.1 & \phn   98.6 &  1.186 &  0.156 & \phs0.260 \\
54984.876 & 0.561 &    270.9 & \phn   97.3 &  1.175 &  0.163 & \phs0.288 \\
54984.882 & 0.561 &    274.5 & \phn   96.1 &  0.926 &  0.124 & \phs0.079 \\
54984.887 & 0.561 &    278.1 & \phn   94.8 &  0.926 &  0.121 & \phs0.122 \\
54984.892 & 0.561 &    281.4 & \phn   93.6 &  0.785 &  0.105 & \phs0.013 \\
54984.904 & 0.561 &    276.2 &       159.5 &  1.145 &  0.151 & \phs0.179 \\
54984.907 & 0.561 &    276.5 &       158.2 &  1.150 &  0.151 & \phs0.194 \\
54984.916 & 0.561 &    276.9 &       156.2 &  1.012 &  0.139 & \phs0.077 \\
54984.921 & 0.561 &    277.2 &       154.9 &  1.078 &  0.151 & \phs0.161 \\
55014.938 & 0.657 &    330.5 &       193.5 &  1.151 &  0.152 & \phs0.188 \\
55014.943 & 0.657 &    330.6 &       192.3 &  0.988 &  0.131 & \phs0.032 \\
55014.948 & 0.657 &    330.6 &       191.3 &  1.023 &  0.137 & \phs0.076 \\
55014.952 & 0.657 &    330.6 &       190.1 &  0.986 &  0.133 & \phs0.051 \\
55014.957 & 0.658 &    330.6 &       188.9 &  1.009 &  0.145 & \phs0.088 \\
55014.963 & 0.658 &    330.7 &       187.4 &  0.819 &  0.144 &  $-$0.081 \\
55014.969 & 0.658 &    330.7 &       186.0 &  0.909 &  0.203 & \phs0.030 \\
55014.975 & 0.658 &    330.7 &       184.6 &  0.790 &  0.106 &  $-$0.066 \\
55014.981 & 0.658 &    330.7 &       183.4 &  0.729 &  0.102 &  $-$0.106 \\
55014.986 & 0.658 &    330.7 &       182.0 &  0.728 &  0.099 &  $-$0.084 \\
55014.992 & 0.658 &    330.7 &       181.0 &  0.765 &  0.103 &  $-$0.031 \\
55054.879 & 0.785 &    248.0 &       178.1 &  1.020 &  0.149 & \phs0.274 \\
55054.883 & 0.785 &    248.0 &       177.0 &  1.245 &  0.182 & \phs0.497 \\
55054.888 & 0.785 &    248.0 &       176.0 &  1.041 &  0.169 & \phs0.289 \\
55054.892 & 0.785 &    247.9 &       174.9 &  0.846 &  0.125 & \phs0.086 \\
55055.855 & 0.788 &    248.0 &       182.7 &  1.144 &  0.173 & \phs0.386 \\
55055.859 & 0.788 &    248.0 &       181.9 &  1.075 &  0.167 & \phs0.322 \\
55055.865 & 0.788 &    248.0 &       180.9 &  1.021 &  0.177 & \phs0.274 \\
55055.895 & 0.789 &    247.9 &       174.0 &  0.603 &  0.116 &  $-$0.190 \\
55055.897 & 0.789 &    247.9 &       173.1 &  0.630 &  0.112 &  $-$0.175 \\
55055.903 & 0.789 &    247.9 &       172.0 &  0.776 &  0.126 &  $-$0.048 \\
55056.827 & 0.792 &    247.8 &       188.6 &  1.337 &  0.189 & \phs0.521 \\
55056.835 & 0.792 &    247.9 &       186.9 &  1.172 &  0.281 & \phs0.384 \\
55056.841 & 0.792 &    247.9 &       185.4 &  1.563 &  0.316 & \phs0.794 \\
55056.847 & 0.792 &    248.0 &       184.2 &  1.135 &  0.199 & \phs0.379 \\
55056.853 & 0.792 &    248.0 &       183.0 &  1.316 &  0.209 & \phs0.567 \\
55056.858 & 0.792 &    248.0 &       181.8 &  1.140 &  0.173 & \phs0.394 \\
55056.867 & 0.792 &    248.0 &       179.3 &  1.084 &  0.168 & \phs0.330 \\
55056.873 & 0.792 &    248.0 &       178.1 &  1.032 &  0.156 & \phs0.267 \\
55056.884 & 0.792 &    248.0 &       175.6 &  0.942 &  0.149 & \phs0.146 \\
55516.641 & 0.263 &    245.7 &       117.9 &  1.069 &  0.166 & \phs0.206 \\
55516.651 & 0.263 &    330.7 &       172.7 &  1.306 &  0.183 & \phs0.555 
\enddata 
\end{deluxetable} 

%Table 4: Spectroscopy 
\clearpage 
\begin{deluxetable}{cccccl} 
\tablewidth{0pc} 
\tabletypesize{\small} 
%\rotate 
\tablenum{4} 
\tablecaption{Journal of Spectroscopy \label{tab4}} 
\tablehead{ 
\colhead{Source} & 
\colhead{Date} & 
\colhead{Range} & 
\colhead{Resolving Power} & 
\colhead{} & 
\colhead{Observatory/Telescope/} \\ 
\colhead{Number} & 
\colhead{(BY)} & 
\colhead{(\AA)} & 
\colhead{($\lambda/\triangle\lambda$)} & 
\colhead{$N$} & 
\colhead{Spectrograph}} 
\startdata 
 1\dotfill & 1995.5 & 5844 -- 5904 &   26000 &   1 & Ritter/1m/Echelle\tablenotemark{a} \\ 
 2\dotfill & 1995.6 & 5572 -- 5895 &   22200 &   1 & KPNO/0.9m/Coud\'{e} \\ 
 3\dotfill & 1995.6 & 6434 -- 6751 &   31000 &   1 & KPNO/0.9m/Coud\'{e} \\ 
 4\dotfill & 1998.7 & 6314 -- 6978 &   12200 &   1 & KPNO/0.9m/Coud\'{e} \\ 
 5\dotfill & 1999.8 & 5401 -- 6735 &\phn5600 &   4 & KPNO/0.9m/Coud\'{e} \\ 
 6\dotfill & 2000.7 & 6443 -- 7108 &   12500 &   1 & KPNO/0.9m/Coud\'{e} \\ 
 7\dotfill & 2001.0 & 6443 -- 7108 &   12500 &   3 & KPNO/0.9m/Coud\'{e}\tablenotemark{b} \\ 
 8\dotfill & 2002.4 & 4692 -- 6018 &\phn4900 &   1 & KPNO/0.9m/Coud\'{e} \\ 
 9\dotfill & 2002.4 & 5980 -- 7313 &\phn6100 &   1 & KPNO/0.9m/Coud\'{e} \\ 
10\dotfill & 2004.7 & 4000 -- 6800 &   34200 &   2 & OHP/1.9m/Elodie\tablenotemark{c} \\ 
11\dotfill & 2004.8 & 6466 -- 7176 &\phn7900 &   1 & KPNO/0.9m/Coud\'{e} \\ 
12\dotfill & 2005.9 & 4236 -- 4587 &   10300 &   2 & KPNO/0.9m/Coud\'{e} \\ 
13\dotfill & 2006.8 & 6466 -- 7176 &\phn7900 &   2 & KPNO/0.9m/Coud\'{e} \\ 
14\dotfill & 2006.8 & 4236 -- 4587 &   10300 &   2 & KPNO/0.9m/Coud\'{e} \\ 
15\dotfill & 2008.6 & 4465 -- 4586 &   76300 &   2 & CFHT/3.6m/ESPaDOnS \\ 
16\dotfill & 2009.9 & 4000 -- 4720 &   75900 &   1 & NOT/2.6m/FIES \\ 
17\dotfill & 2010.5 & 3994 -- 4663 &\phn5700 &   2 & KPNO/4.0m/R-C \\ 
18\dotfill & 2010.5 & 3873 -- 4540 &\phn6400 &   1 & Lowell/1.8m/DeVeny \\ 
19\dotfill & 2010.6 & 4292 -- 4670 &\phn4300 &   1 & DAO/1.8m/Cassegrain \\ 
20\dotfill & 2010.6 & 3873 -- 4540 &\phn6400 &   1 & Lowell/1.8m/DeVeny  
\enddata 
\tablenotetext{a}{http://astro1.panet.utoledo.edu/$^\sim$wwritter/archive/PREST-archive.html}
\tablenotetext{b}{http://www.noao.edu/cflib/}
\tablenotetext{c}{http://atlas.obs-hp.fr/elodie/}
\end{deluxetable} 
 
%Table 5: radial velocities for Ab1
\begin{deluxetable}{cccccccc}
\tabletypesize{\scriptsize}
\tablewidth{0pt}
\tablenum{5}
\tablecaption{Radial Velocity Measurements for Ab1\label{tab5}}
\tablehead{
\colhead{Date}          &
\colhead{$\phi$}        &
\colhead{$\phi$}	&
\colhead{$V_{r}$}	&
\colhead{$\sigma (V_{r})$}&
\colhead{$\Delta V_{r}$(blend)}&
\colhead{$O-C$}         &
\colhead{Source} \\
\colhead{(HJD$-$2,400,000)}&
\colhead{(close)}       &
\colhead{(wide)}        &
\colhead{(km s$^{-1}$)} &
\colhead{(km s$^{-1}$)} &
\colhead{(km s$^{-1}$)} &
\colhead{(km s$^{-1}$)} &
\colhead{Number\tablenotemark{a}}	
}
\startdata
22815.476 &  0.588 &  0.966 & 
          $ -16.1$ & 8.0 & \phn     $  -2.3$ &\phn     $  -4.6$ &\phn  0 \\
23952.434 &  0.227 &  0.054 & 
 \phn\phs $   3.1$ & 8.0 & \phn\phs $   1.7$ &\phs     $  10.4$ &\phn  0 \\
23962.408 &  0.259 &  0.055 & 
 \phn\phs $   3.8$ & 8.0 & \phn\phs $   1.5$ &\phs     $  15.2$ &\phn  0 \\
24254.017 &  0.192 &  0.078 & 
 \phn\phs $   8.8$ & 8.0 & \phn\phs $   2.3$ &\phs     $  14.1$ &\phn  0 \\
24424.648 &  0.739 &  0.091 & 
          $ -17.1$ & 8.0 & \phn     $  -2.3$ &\phn     $  -6.8$ &\phn  0 \\
24668.949\tablenotemark{b} &  0.521 &  0.110 & 
 \phn\phs $   1.1$ & 8.0 & \phn     $  -0.5$ &\phs     $  33.2$ &\phn  0 \\
24673.884\tablenotemark{b} &  0.536 &  0.110 & 
 \phn     $  -6.3$ & 8.0 & \phn     $  -2.7$ &\phs     $  23.3$ &\phn  0 \\
24675.927 &  0.543 &  0.111 & 
          $ -17.0$ & 8.0 & \phn     $  -3.1$ &\phs     $  11.9$ &\phn  0 \\
39002.424 &  0.402 &  0.225 & 
          $ -26.0$ & 8.0 &          $ -15.9$ &         $ -13.5$ &\phn  0 \\
40044.910 &  0.739 &  0.306 & 
          $ -14.9$ & 1.4 & \phn     $  -1.0$ &\phn     $  -4.6$ &\phn  0 \\
40065.488 &  0.805 &  0.308 & 
 \phn\phs $   0.0$ & 8.0 & \phn\phs $   3.1$ &\phn\phs $   5.7$ &\phn  0 \\
40347.870 &  0.708 &  0.330 & 
          $ -11.0$ & 1.1 & \phn     $  -1.0$ &\phn\phs $   2.9$ &\phn  0 \\
43741.479 &  0.571 &  0.594 & 
          $ -21.5$ & 1.3 & \phn     $  -0.0$ &\phn\phs $   1.7$ &\phn  0 \\
43741.516 &  0.571 &  0.594 & 
          $ -21.7$ & 1.3 & \phn     $  -0.0$ &\phn\phs $   1.5$ &\phn  0 \\
43771.710 &  0.668 &  0.596 & 
          $ -17.7$ & 2.1 & \phn     $  -1.5$ &\phn     $  -4.7$ &\phn  0 \\
43772.750 &  0.671 &  0.596 & 
          $ -16.4$ & 3.1 & \phn     $  -1.4$ &\phn     $  -3.7$ &\phn  0 \\
43777.740 &  0.687 &  0.596 & 
 \phn     $  -8.1$ & 3.1 & \phn     $  -0.4$ &\phn\phs $   3.7$ &\phn  0 \\
44051.110 &  0.562 &  0.618 & 
          $ -21.9$ & 1.3 & \phn     $  -0.0$ &\phn\phs $   1.2$ &\phn  0 \\
44087.791 &  0.680 &  0.621 & 
          $ -11.8$ & 1.3 & \phn\phs $   0.0$ &\phn\phs $   0.8$ &\phn  0 \\
44593.740 &  0.299 &  0.660 & 
 \phn     $  -4.5$ & 1.3 & \phn\phs $   0.4$ &\phn\phs $   4.9$ &\phn  0 \\
45659.984 &  0.712 &  0.743 & 
 \phn     $  -0.2$ & 1.3 & \phn\phs $   0.3$ &\phn\phs $   5.3$ &\phn  0 \\
45991.853 &  0.775 &  0.769 & 
 \phn\phs $   8.6$ & 1.3 & \phn\phs $   0.6$ &\phn\phs $   5.4$ &\phn  0 \\
46606.058 &  0.741 &  0.816 & 
 \phn     $  -3.2$ & 3.3 & \phn\phs $   0.5$ &\phn     $  -3.5$ &\phn  0 \\
46607.015 &  0.744 &  0.816 & 
 \phn     $  -1.9$ & 2.2 & \phn\phs $   0.6$ &\phn     $  -2.6$ &\phn  0 \\
46608.051 &  0.747 &  0.817 & 
 \phn     $  -2.4$ & 2.1 & \phn\phs $   0.6$ &\phn     $  -3.6$ &\phn  0 \\
46609.062 &  0.750 &  0.817 & 
 \phn     $  -0.9$ & 1.1 & \phn\phs $   0.6$ &\phn     $  -2.4$ &\phn  0 \\
46612.036 &  0.760 &  0.817 & 
 \phn\phs $   3.1$ & 1.4 & \phn\phs $   0.8$ &\phn\phs $   0.5$ &\phn  0 \\
46985.281 &  0.955 &  0.846 & 
 \phs     $  25.0$ & 1.3 & \phn\phs $   0.4$ &\phn\phs $   2.0$ &\phn  0 \\
46986.266 &  0.958 &  0.846 & 
 \phs     $  21.5$ & 1.3 & \phn\phs $   0.4$ &\phn     $  -1.6$ &\phn  0 \\
46986.660 &  0.959 &  0.846 & 
 \phs     $  23.2$ & 1.3 & \phn\phs $   0.4$ &\phn\phs $   0.0$ &\phn  0 \\
46986.707 &  0.959 &  0.846 & 
 \phs     $  21.1$ & 1.3 & \phn\phs $   0.4$ &\phn     $  -2.1$ &\phn  0 \\
46986.778 &  0.959 &  0.846 & 
 \phs     $  22.4$ & 1.3 & \phn\phs $   0.4$ &\phn     $  -0.8$ &\phn  0 \\
46988.682 &  0.966 &  0.846 & 
 \phs     $  23.3$ & 1.3 & \phn\phs $   0.4$ &\phn     $  -0.1$ &\phn  0 \\
46988.724 &  0.966 &  0.846 & 
 \phs     $  24.2$ & 1.3 & \phn\phs $   0.4$ &\phn\phs $   0.8$ &\phn  0 \\
46988.769 &  0.966 &  0.846 & 
 \phs     $  21.6$ & 1.3 & \phn\phs $   0.4$ &\phn     $  -1.8$ &\phn  0 \\
46988.815 &  0.966 &  0.846 & 
 \phs     $  22.6$ & 1.3 & \phn\phs $   0.4$ &\phn     $  -0.8$ &\phn  0 \\
47773.924 &  0.479 &  0.907 & 
          $ -14.1$ & 1.3 & \phn     $  -0.0$ &\phn\phs $   1.6$ &\phn  0 \\
47773.969 &  0.479 &  0.907 & 
          $ -15.0$ & 1.3 & \phn     $  -0.0$ &\phn\phs $   0.7$ &\phn  0 \\
49231.714 &  0.145 &  0.021 & 
 \phs     $  10.9$ & 2.9 & \phn\phs $   5.2$ &\phn\phs $   7.1$ &\phn  0 \\
49236.773 &  0.162 &  0.021 & 
 \phn\phs $   0.0$ & 2.1 & \phn\phs $   2.0$ &\phn     $  -5.1$ &\phn  0 \\
49614.855 &  0.372 &  0.050 & 
          $ -24.7$ & 5.4 & \phn\phs $   0.5$ &\phn     $  -1.4$ &\phn  0 \\
49840.255 &  0.093 &  0.068 & 
 \phn\phs $   8.5$ & 1.3 & \phn\phs $   0.2$ &\phn\phs $   0.9$ &\phn  0 \\
49842.803 &  0.102 &  0.068 & 
 \phn\phs $   5.8$ & 4.0 & \phn\phs $   8.0$ &\phn\phs $   6.7$ &\phn  0 \\
49843.780 &  0.105 &  0.068 & 
 \phs     $  10.4$ & 4.2 & \phs     $  10.1$ &\phs     $  13.6$ &\phn  0 \\
49916.748 &  0.338 &  0.074 & 
          $ -23.1$ & 5.1 & \phn     $  -0.4$ &\phn     $  -2.1$ &\phn  1 \\
49942.734 &  0.421 &  0.076 & 
          $ -30.0$ & 0.8 & \phn     $  -0.6$ &\phn     $  -1.6$ &\phn  2 \\
49942.886 &  0.422 &  0.076 & 
          $ -23.9$ & 0.9 & \phn     $  -6.8$ &\phn     $  -1.6$ &\phn  3 \\
49982.849 &  0.550 &  0.079 & 
          $ -28.0$ & 1.3 & \phn\phs $   0.2$ &\phn\phs $   2.8$ &\phn  0 \\
49985.625 &  0.559 &  0.079 & 
          $ -16.1$ & 2.9 &          $ -18.7$ &\phn     $  -4.5$ &\phn  0 \\
50059.437\tablenotemark{b} &  0.795 &  0.085 & 
 \phs     $  16.3$ & 6.5 & \phs     $  11.7$ &\phs     $  33.0$ &\phn  0 \\
51056.748 &  0.987 &  0.163 & 
 \phs     $  10.3$ & 1.9 & \phn\phs $   5.0$ &\phn\phs $   6.1$ &\phn  4 \\
51466.709 &  0.300 &  0.195 & 
          $ -17.3$ & 2.0 & \phn     $  -1.4$ &\phn     $  -0.6$ &\phn  5 \\
51467.788 &  0.303 &  0.195 & 
          $ -12.3$ & 2.0 & \phn     $  -1.2$ &\phn\phs $   5.0$ &\phn  5 \\
51467.795 &  0.303 &  0.195 & 
          $ -12.1$ & 2.0 & \phn     $  -1.2$ &\phn\phs $   5.2$ &\phn  5 \\
51468.754 &  0.306 &  0.195 & 
          $ -17.8$ & 2.2 & \phn     $  -1.5$ &\phn     $  -0.4$ &\phn  5 \\
51817.657 &  0.423 &  0.222 & 
          $ -29.3$ & 1.4 & \phn     $  -3.4$ &\phn     $  -2.8$ &\phn  6 \\
51888.616 &  0.650 &  0.227 & 
          $ -16.8$ & 1.2 & \phn     $  -1.7$ &\phn\phs $   5.1$ &\phn  7 \\
51893.557 &  0.666 &  0.228 & 
          $ -14.1$ & 2.5 & \phn     $  -1.3$ &\phn\phs $   6.4$ &\phn  7 \\
51895.593 &  0.672 &  0.228 & 
          $ -11.7$ & 2.9 & \phn     $  -1.0$ &\phn\phs $   8.3$ &\phn  7 \\
52430.950\tablenotemark{b} &  0.386 &  0.270 & 
          $ -11.3$ & 2.8 & \phn     $  -2.2$ &\phs     $  12.8$ &\phn  8 \\
52436.914 &  0.405 &  0.270 & 
          $ -26.8$ & 9.5 & \phn     $  -5.7$ &\phn     $  -4.5$ &\phn  9 \\
53246.460 &  0.997 &  0.333 & 
 \phn\phs $   8.2$ & 0.9 & \phn\phs $   1.3$ &\phn     $  -2.3$ &     10 \\
53247.481 &  1.000 &  0.333 & 
 \phn\phs $   8.4$ & 1.1 & \phn\phs $   1.4$ &\phn     $  -2.0$ &     10 \\
53290.656 &  0.138 &  0.336 & 
 \phn     $  -1.8$ & 2.3 & \phn\phs $   0.9$ &\phn     $  -5.3$ &     11 \\
53683.603 &  0.396 &  0.367 & 
          $ -21.4$ & 1.0 & \phn     $  -2.4$ &\phn\phs $   1.6$ &     12 \\
53684.593 &  0.399 &  0.367 & 
          $ -21.3$ & 1.0 & \phn     $  -2.4$ &\phn\phs $   2.1$ &     12 \\
54019.652 &  0.472 &  0.393 & 
          $ -30.1$ & 2.3 & \phn     $  -3.4$ &\phn     $  -4.4$ &     13 \\
54024.715 &  0.488 &  0.394 & 
          $ -26.9$ & 2.3 & \phn     $  -3.1$ &\phn     $  -0.7$ &     13 \\
54029.707 &  0.504 &  0.394 & 
          $ -25.9$ & 0.9 & \phn     $  -3.1$ &\phn\phs $   0.3$ &     14 \\
54031.627 &  0.510 &  0.394 & 
          $ -23.9$ & 1.0 & \phn     $  -2.9$ &\phn\phs $   2.5$ &     14 \\
54675.892 &  0.572 &  0.444 & 
          $ -27.3$ & 0.7 & \phn     $  -2.2$ &\phn     $  -3.2$ &     15 \\
54675.917 &  0.572 &  0.444 & 
          $ -27.3$ & 0.6 & \phn     $  -2.2$ &\phn     $  -3.2$ &     15 \\
55146.400 &  0.078 &  0.481 & 
 \phn\phs $   9.8$ & 0.2 & \phn\phs $   2.2$ &\phn     $  -0.2$ &     16 \\
55366.903 &  0.784 &  0.498 & 
 \phn\phs $   7.3$ & 2.5 & \phn\phs $   2.6$ &\phs     $  11.7$ &     17 \\
55369.913 &  0.794 &  0.498 & 
 \phn\phs $   2.6$ & 2.5 & \phn\phs $   1.9$ &\phn\phs $   4.9$ &     17 \\
55383.939 &  0.839 &  0.499 & 
 \phn\phs $   6.9$ & 1.3 & \phn\phs $   3.4$ &\phn\phs $   5.3$ &     18 \\
55402.849 &  0.899 &  0.501 & 
 \phn\phs $   4.2$ & 2.4 & \phn\phs $   5.3$ &\phn     $  -1.3$ &     19 \\
55402.871 &  0.899 &  0.501 & 
 \phn\phs $   3.1$ & 1.4 & \phn\phs $   2.9$ &\phn     $  -4.9$ &     20 
\enddata
\tablenotetext{a}{0: McKibben et al.\ (1998); 1--20: see Table~4.}
\tablenotetext{b}{Assigned zero weight in the orbital solution.}
\end{deluxetable}

%Table 6: Orbital Elements
\begin{deluxetable}{lcccc}
\tabletypesize{\scriptsize}
\tablewidth{0pc}
\tablenum{6}
\tablecaption{Radial Velocity Orbital Elements for Ab1\label{tab6}}
\tablehead{
\colhead{Orbital} &
\colhead{Aa,Ab System} &
\colhead{Aa,Ab System} &
\colhead{Ab1,Ab2 System} &
\colhead{Ab1,Ab2 System} \\
\colhead{Element} &
\colhead{(no correction)} &
\colhead{(blend correction)} &
\colhead{(no correction)} &
\colhead{(blend correction)} 
}
\startdata
$P$~(y)                 \dotfill & $35.20$\tablenotemark{a} & $35.20$\tablenotemark{a} 
                                 & $0.85543 \pm 0.00030$    & $0.85533 \pm 0.00029$ \\
$P$~(d)                 \dotfill & $12855$\tablenotemark{a} & $12855$\tablenotemark{a} 
                                 & $312.44 \pm 0.11$        & $312.40 \pm 0.10$     \\
$T$ (BY)                \dotfill & $1993.1 \pm 0.4$         & $1992.9 \pm 0.3$         
                                 & $1996.111 \pm 0.005$     & $1996.109 \pm 0.004$  \\
$T$ (HJD--2,400,000)    \dotfill & $49030 \pm 148$          & $48966 \pm 103$          
                                 & $50124.1 \pm 1.7$        & $50123.5 \pm 1.5$     \\
$K$ (km s$^{-1}$)       \dotfill & $8.6 \pm 0.6$            & $8.7 \pm 0.4$            
                                 & $19.2 \pm 0.4$           & $21.1 \pm 0.4$        \\
$\gamma$ (km s$^{-1}$)  \dotfill & $-4.3 \pm 0.4$           & $-4.7 \pm 0.4$            
                                 & $-4.3 \pm 0.4$           & $-4.7 \pm 0.4$        \\
$e$                     \dotfill & $0.489$\tablenotemark{a} & $0.489$\tablenotemark{a} 
                                 & $0.0$                    & $0.0$                 \\
$\omega$ (deg)          \dotfill & $70.4$\tablenotemark{a}  & $70.4$\tablenotemark{a}  
                                 & \nodata                  & \nodata               \\
$f(m)$ ($M_\odot$)      \dotfill & $0.56 \pm 0.12$          & $0.58 \pm 0.08$          
                                 & $0.230 \pm 0.015$        & $0.306 \pm 0.019$     \\
$a_1\sin i$ ($10^6$ km) \dotfill & $1327 \pm 94$            & $1341 \pm 65$            
                                 & $82.5 \pm 1.8$           & $90.8 \pm 1.9$        \\
rms (km s$^{-1}$)       \dotfill & 3.0                      & 3.1                      
                                 & 3.0                      & 3.1                 
\enddata
\tablenotetext{a}{Fixed with values from the visual orbit (Table 2).}
\end{deluxetable}

%Table 7: radial velocities for Aa
\begin{deluxetable}{cccccl}
%\tabletypesize{\scriptsize}
\tablewidth{0pt}
\tablenum{7}
\tablecaption{Radial Velocity Measurements for Aa\label{tab7}}
\tablehead{
\colhead{Date}          &
\colhead{$\phi$}        &
\colhead{$V_{r}$}	&
\colhead{$O-C$}         &
\colhead{}              &
\colhead{}              \\
\colhead{(HJD$-$2,400,000)}&
\colhead{(wide)}        &
\colhead{(km s$^{-1}$)} &
\colhead{(km s$^{-1}$)} &
\colhead{S/N}           &
\colhead{Source}	}
\startdata
 46608.445 &  0.817 &     $-35.0$ &\phn     $ -6.3$ &\phn 960 &  CFHT/1986 \\
 49942.734 &  0.076 &     $-39.5$ &         $-40.9$ &\phn 380 &  KPNO/1995 \\
 51467.762 &  0.195 &     $-24.6$ &         $-28.9$ &    1020 &  KPNO/1999 \\
 53246.970 &  0.333 &\phn $ -0.7$ &\phs\phn $  0.3$ &\phn 240 &  OHP/2004  \\
 54675.904 &  0.444 &\phn $ -2.3$ &\phs\phn $  3.8$ &\phn 750 &  CFHT/2008 \\
 55146.400 &  0.481 &\phs $ 23.2$ &\phs     $ 31.1$ &\phn 220 &  NOT/2009  \\
\enddata
\end{deluxetable}

%Table 8
\begin{deluxetable}{lcccc}
\tablewidth{0pc}
\tablenum{8}
\tablecaption{Representative Stellar Parameters\label{tab8}}
\tablehead{
\colhead{Parameter} &
\colhead{Aa} &
\colhead{Ab1} &
\colhead{Ab2} &
\colhead{B}}
\startdata
$F/F_{\rm total}$        \dotfill              &   0.43     &  0.40     &  0.06    & 0.11 \\
$T_{\rm eff}$ (kK)       \dotfill              &   33       &  32.5     &  20      & 23   \\
$\log g$ (cm s$^{-2}$)   \dotfill              &    4.0     &   3.5     &  4.0     & 4.0  \\
$V \sin i$ (km s$^{-1}$) \dotfill              &   350      &   40      &  200     & 100  \\
$M$ ($M_\odot$)          \dotfill              &    21      &   23      &    9     & \nodata  \\
$M_V$ (mag)              \dotfill              &  $-3.6$    & $-3.5$    & $-1.4$   & $-2.1$ 
\enddata
\end{deluxetable}

%%%%%%%%%%%%%%%%%%%%%%%%%%%%%%%%%%%%%%%%%%%%%%%%%%%%%%%%%%%%%%

% Figure captions
\clearpage

\input{epsf}

% Figure 1

% Figure 2

% Figure 3 

% Figure 4

% Figure 5

% Figure 6

% Figure 7

% Figure 8

% Figure 9

% Figure 10

% Figure 11

%%%%%%%%%%%%%%%%%%%%%%%%%%%%%%%%%%%%%%%%%%%%%%%%%%%%%%%%%%%%%%%

\end{document}